\documentclass{article}

\usepackage{arxiv}

\usepackage[utf8]{inputenc} 
\usepackage[T1]{fontenc}    
\usepackage{hyperref}       
\usepackage{url}            
\usepackage{booktabs}       
\usepackage{amsfonts}       
\usepackage{nicefrac}       
\usepackage{microtype}      
\usepackage{lipsum}
\usepackage{graphicx}

\usepackage{amsmath,amssymb,amsfonts}
\usepackage{algorithm}
\usepackage{algpseudocode}
\usepackage{graphicx}
\usepackage{caption}
\usepackage{subcaption}
\usepackage{textcomp}

\usepackage{multirow}
\usepackage{makecell}
\usepackage[group-separator={,}]{siunitx}
\usepackage{hyperref}

\title{OpSparse: a Highly Optimized Framework for Sparse General Matrix Multiplication on GPUs}

\author{
Zhaoyang~Du\thanks{This work has been submitted to the IEEE for possible publication. Copyright may be transferred without notice, after which this version may no longer be accessible. This work has been submitted to the IEEE Access since May 7, 2022, and is currently under review.}\\
 College of  Information  Science  and  Electronic Engineering\\
Zhejiang University\\
Hangzhou 310007, China \\
	\texttt{11731021@zju.edu.cn} \\
	\And
	Yijin~Guan \\
	Alibaba Group\\
	Hangzhou 311121, China\\
	\And
    Tianchan~Guan\\
    Alibaba Group\\
	Hangzhou 311121, China\\
	\And
	Dimin~Niu\\
	Alibaba Group\\
	Hangzhou 311121, China\\
	\And
    Linyong~Huang\\
    College of  Information  Science  and  Electronic Engineering\\
    Zhejiang University\\
    Hangzhou 310007, China \\
	\And
	Hongzhong~Zheng\\
	Alibaba Group\\
	Hangzhou 311121, China\\
	\And
	Yuan~Xie\\
    Alibaba Group\\
	Hangzhou 311121, China	\\
}

\begin{document}
\maketitle
\begin{abstract}
Sparse general matrix multiplication (SpGEMM) is an important and expensive computation primitive in many real-world applications. Due to SpGEMM's inherent irregularity and the vast diversity of its input matrices, developing high-performance SpGEMM implementation on modern processors such as GPUs is challenging. The state-of-the-art SpGEMM libraries (i.e., $nsparse$ and $spECK$) adopt several algorithms to tackle the challenges of global load balance, local load balance, and allocation of the result matrix. While these libraries focus on the high-level algorithm design for SpGEMM, they neglect several low-level architecture-specific optimizations, which causes inefficient implementations in their libraries. In this paper, we classify their inefficient implementations into seven categories. Based on our observations, we propose a highly optimized SpGEMM library called $OpSparse$. The optimizations in $OpSparse$ include 1) optimizing the binning method by improving the utilization of the shared memory, 2) optimizing the hashing method by reducing the access to the hash table, 3) improving the trade-off between hash collision rate and hardware utilization in the hashing method by setting appropriate binning ranges, 4) reducing the overheads of global memory utilization by minimizing the global memory usage of the metadata, and 5) improving the execution parallelism by overlapping global memory allocation with kernel execution. Performance evaluations with 26 commonly used matrices on an Nvidia Tesla V100 GPU show that $OpSparse$ achieves up to $27.8\times$, $1.81\times$, and $2.04\times$ performance speedup over three state-of-the-art libraries: $cuSPARSE$, $nsparse$, and $spECK$, respectively.
\end{abstract}


\section{Introduction}\label{sec:intro}
Sparse general matrix multiplication (SpGEMM) is widely used in many real-world applications such as algebraic multigrid solvers~\cite{AMG, AMG2}, Markov clustering~\cite{markov}, multi-source breadth first search~\cite{BFS}, molecular dynamics simulations~\cite{molecular}, and finite element simulations based on domain decomposition~\cite{finite-element}.

The extensive use of SpGEMM in real-world applications has led to the development of several SpGEMM libraries on CPUs~\cite{matlab, yusuke, kokkos, pb-SpGEMM}, GPUs~\cite{c_cusparse, c_bhsparse, c_nsparse, c_speck}, and accelerators~\cite{outerspace, sparch, matraptor, gamma}, targeting high-performance computing. This paper focuses on developing a high-performance SpGEMM library on GPUs.

The inherent irregularity and the vast diversity of the input matrices cause many challenges that hinder the performance of SpGEMM on GPUs. Irregularity is introduced by the condensed storage format used by the input and output matrices of SpGEMM, which causes challenges including dataflow selection, accumulator algorithm design, and load imbalance among sub-tasks. In addition, the hard-to-predict sparse structure of the output matrix makes its memory allocation in condensed storage format challenging, which further affects the performance of SpGEMM.

The state-of-the-art SpGEMM libraries (i.e., $nsparse$~\cite{c_nsparse} and $spECK$~\cite{c_speck}) adopt several algorithms to tackle the challenges of global load balance, local load balance, and memory allocation of the result matrix. While these libraries focus on the high-level algorithm design for SpGEMM, they neglect several low-level architecture-specific optimizations, which cause inefficient implementations in their libraries. For example, the excessive global memory accesses in the binning method, the excessive accesses to the hash tables in the hashing method, and the excessive global memory allocation for the metadata are inefficient implementations in their libraries. Moreover, the parallelism of global memory allocation and kernel execution, which can improve the performance of SpGEMM, is not utilized in these SpGEMM libraries.

In this paper, we identify seven kinds of inefficient implementations of two state-of-the-art SpGEMM libraries, i.e., $nsparse$ and $spECK$. Then we propose optimizations for each of the inefficient implementations and integrate them into the proposed framework named $OpSparse$. According to our experiments on diverse sparse matrices, $OpSparse$ achieves significant performance improvement over the libraries $nsparse$ and $spECK$. We also conduct additional experiments to show the performance improvements of several individual optimizations to provide insights of the low-level architecture-specific optimizations of SpGEMM.

The key technical contributions of this work are as follows:
\begin{itemize}
    \item We identify seven kinds of inefficient architectural implementations of two state-of-the-art SpGEMM libraries.
    \item We optimize all the identified inefficient implementations and propose a library named $OpSparse$, integrating all the optimizations we developed. 
    \item We evaluate the performance of the proposed $OpSparse$ framework on 26 commonly used benchmarks. The results show that the proposed $OpSparse$ achieves on average $7.35\times$ (up to $27.8\times$), $1.43\times$ (up to $1.81\times$), and $1.52\times$ (up to $2.04\times$) speedups over three SpGEMM libraries $cuSPARSE$, $nsparse$, and $spECK$, respectively.
\end{itemize}

The rest of the paper is organized as follows. Section~\ref{sec:bg} and section~\ref{sec:related} provide background and related work, respectively. Section~\ref{sec:motivation} describes the inefficient implementations in the existing libraries. Section~\ref{sec:method} describes the proposed SpGEMM framework ($OpSparse$) and the corresponding optimizations. Section~\ref{sec:evaluate} first shows the overall performance compared to three state-of-the-art libraries; then, it shows the performance improvements of several individual optimizations as compared to the baseline. Section~\ref{sec:conclusion} concludes this paper.

\section{Background}\label{sec:bg}
In this section, we provide the background of SpGEMM algorithm on GPUs.

\subsection{Row-wise SpGEMM}\label{sec:row-wise}
In SpGEMM, given two sparse input matrices $A$ and $B$, each element of the output matrix $C$ is computed as:
\begin{equation}C_{ij} = \sum_{k}A_{ik}\cdot{B}_{kj},\label{eq:inner}\end{equation}
where $i$ and $j$ are the row and column indices of the nonzero elements of $A$ and $B$, respectively, and $k$ is the set of the colliding indices. 

One variation of the above computation pattern of SpGEMM is :
\begin{equation}C_{i*} = \sum_{k}A_{ik}\cdot{B}_{k*},\label{eq:row-wise}\end{equation}
where $C_{i*}$ and $A_{i*}$ represent all the nonzero elements in the $i_{th}$ row of $C$ and $A$, respectively, $k$ belongs to the set of column indices of the nonzero elements in the ith row of $A$, and $B_{k*}$ represents all the nonzero elements in the $k_{th}$ row of $B$.

Equation~(\ref{eq:row-wise}) describes the row-wise dataflow to perform SpGEMM. The row-wise dataflow is commonly adopted in the state-of-the-art SpGEMM libraries~\cite{c_cusparse, c_bhsparse, c_nsparse, c_speck}. There are three key benefits of the row-wise dataflow: 1) zero elements are completely avoided in both computation and memory access, 2) the computation of each output row is independent of each other; therefore, the row-wise SpGEMM can be easily parallelized, and 3) accumulating the intermediate products of each output row has good temporal locality.

\subsubsection{CSR storage format}\label{sec:CSR}
The compressed sparse row (CSR) storage format is one of the most commonly used sparse storage formats for SpGEMM, which is adopted by state-of-the-art SpGEMM libraries such as $cuSPARSE$~\cite{c_cusparse}, $nsparse$~\cite{c_nsparse}, and $spECK$~\cite{c_speck}. In this paper, we also choose CSR as the storage format of the input and output matrices in the proposed SpGEMM framework.

\begin{figure}[h]
\centering
\includegraphics[width=0.45\textwidth]{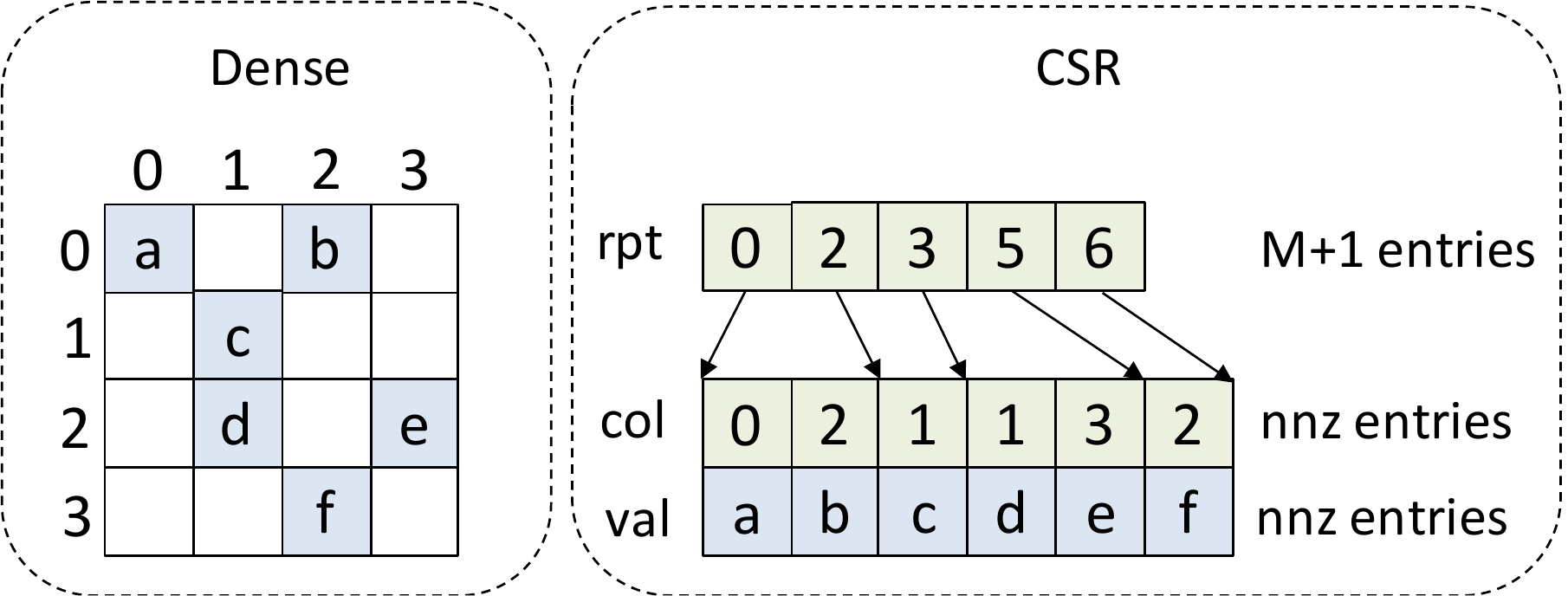}
\caption{Illustration of the CSR storage format. Left: matrix A in the dense storage format. Right: matrix A in the CSR storage format}
\label{fig:CSR}
\end{figure}

Fig.~\ref{fig:CSR} illustrates the CSR storage format. The CSR consists of three arrays to record the nonzero elements and their corresponding indices. The $val$ and $col$ array record the nonzero elements and their corresponding column indices in a sorted row-major and column-major order. The lengths of $val$ and $col$ arrays are both the number of nonzero elements ($n_{nz}$) of the sparse matrix. The $rpt$ array records the start and end positions for each row's values and column indices in the $val$ and $col$ arrays. Since the $i_{th}$ row's end positions can be encoded the same as the $(i+1)_{th}$ row's start positions in the $rpt$ array, the $rpt$ array is further compressed to $M + 1$ entries, where $M$ is the number of rows of the matrix. One of the key performance benefits of using CSR is that it is easy to access the elements of an entire row.

\subsubsection{Compression ratio (CR)}\label{sec:CR}
In row-wise SpGEMM, each output row is computed by accumulating multiple intermediate products. The compression ratio represents the average number of intermediate products for one nonzero element in the result matrix. Therefore, the compression ratio can be calculated by dividing the total number of intermediate products ($n_{prod}$) in performing row-wise SpGEMM by the result matrix's total number of nonzero elements ($n_{nz}$) (Equation~(\ref{eq:CR})). 

\begin{equation}Compression\ ratio = \frac{Total\ n_{prod}\ to\ compute\ C}{Total\ n_{nz}\ of\ C}.\label{eq:CR}\end{equation}

\subsection{One-phase SpGEMM and two-phase SpGEMM}\label{sec:two-phase}
One of the critical issues in row-wise SpGEMM design is that the number of nonzero elements of the result matrix is not known before calculating the result matrix. To tackle such an issue, both one-phase~\cite{c_bhsparse} and two-phase~\cite{c_cusparse, kokkos, c_nsparse, c_speck} methods are used. 

To avoid ambiguity, we only discuss the one-phase and two-phase methods when applying the row-wise dataflow and the CSR storage format in SpGEMM. The two-phase method consists of the symbolic and numeric phases, in which the numeric phase depends upon the result from the symbolic phase. In the symbolic phase, the row size (number of nonzero elements of a row) of each output row is computed with the indices information of the two input matrices. Multiplication operation is avoided in the symbolic phase. After the symbolic phase, the total number of nonzero elements of the result matrix is computed based on the row sizes. Then the memory for the column indices and values of the result matrix is allocated. In the numeric phase, each output row's column indices and values are calculated and stored in the allocated memory space.

In contrast, the one-phase method computes the row sizes, column indices, and values of the result matrix simultaneously. However, due to the unknown memory size of the result matrix, the memory space for storing the column indices and values of the result matrix cannot be precisely allocated in advance. As a result, temporary memory space is required in the one-phase method. The size of the temporary memory space is defined by either estimating the upper-bound row sizes of the result matrix~\cite{c_bhsparse} or using dynamic allocation when needed~\cite{matlab}. The former may result in over-allocation, and the latter is not feasible for GPUs~\cite{kokkos} (due to the high cost-to-benefit ratio of allocating small-footprint memory in GPUs). Moreover, the one-phase method additionally requires two time-consuming operations: 1) allocating the temporary memory space to store the result matrix, and 2) copying the result matrix from the temporary memory space to the standard CSR memory space.

\section{Related work}\label{sec:related}
Researchers have proposed many approaches to improve the performance of SpGEMM on GPUs. We introduce several approaches that use the row-wise and two-phase methods to perform SpGEMM on GPUs.

Demouth proposed the $cuSPARSE$~\cite{c_cusparse} library, which uses the two-phase method to implement SpGEMM. The memory usage for the $C$ matrix is efficient. However, $cuSPARSE$ use the naive load balance method for the computation. They compute all the output rows in the same way (one kernel for the symbolic phase and one kernel for the numeric phase) despite the varying $n_{prod}$ (or $n_{nz}$) per output row. Consequently, their implementation may suffer from severe load imbalance issues. $cuSPARSE$ uses hashing method to compute the result row. However, the implementation of their hashing method is inefficient. Specifically,  because $cuSPARSE$ computes all the output rows with the same kernel, the kernel implements both shared memory and global memory hash tables to compute the rows with varying $n_{prod}$ (or $n_{nz}$). Therefore, the $key$s can first be inserted into the shared memory hash table for fast access speed. However, when insertion to the shared memory hash table fails, it switches to the global memory hash table and recomputes the failed row. This implementation causes recomputations of several rows and potentially inefficient utilization of the hardware resources for the shared memory hash tables.

Nagasaka \emph{et al.} proposed $nsparse$~\cite{c_nsparse}, which improves the global load balance issue by using multiple kernels to compute the output rows with varying $n_{prod}$ (or $n_{nz}$). Specifically, $nsparse$ first groups rows into multiple bins by their $n_{prod}$ (or $n_{nz}$), then it computes the rows in each bin with a more appropriate kernel. For better thread utilization, combines two thread assignment methods. When computing tiny rows, $nsparse$ uses a sub-warp instead of a whole thread block to compute each row and extracts an entire row in B by one thread. When computing other rows, $nsparse$ uses a thread block to compute each row and extracts an entire row in B by a warp (32 threads). In general, $nsparse$ achieved high performance on many commonly used benchmarks and showed an efficient framework for SpGEMM on GPUs. However, there are still several inefficient implementations in $nsparse$'s library, limiting its performance. We identify and optimize them in section~\ref{sec:motivation} to section~\ref{sec:method}.

Parger \emph{et al.} proposed $spECK$~\cite{c_speck} to improve the performance of the two-phase SpGEMM. In addition to the hashing accumulator, $spECK$ also uses the dense accumulator and the single-row accumulator to improve the performance when computing particular rows. For example, it uses the dense accumulator to compute rows with extremely large $n_{nz}$. Moreover, $spECK$ introduces the local load balance method by which the number of threads to extract the rows in B is dynamically determined. To guide their global and local load balance and selection of accumulator methods, $spECK$ introduces an additional lightweight row analysis on the two input matrices. In this paper, we do not adopt the local load balance method and the multiple accumulator methods. The performance improvement of our optimizations is orthogonal to the performance improvement of adopting these two methods. 

\section{Inefficient implementations in the existing libraries}\label{sec:motivation}
There are multiple inefficient implementations in the state-of-the-art SpGEMM libraries, limiting their performance nontrivially. We classify these inefficient implementations into seven categories and describe them in detail. 

\subsection{Binning method}\label{sec:moti-binning-method}
The binning method implements the global load balance for the symbolic and the numeric phases~\cite{c_nsparse, c_speck}. The primary function of the binning method is to classify rows into different bins/groups. One important task of the binning method is to count the $bin\_size$ by the atomic operation. However, existing libraries perform massive atomic operations directly on the global memory, which is inefficient. Considering the cost of the binning steps, our performance profiling shows that the execution time of the binning method in $nsparse$ and $spECK$ takes more than 10\% of the overall execution time in many benchmarks (shown in Fig.~\ref{fig:binning_percent} in section~\ref{sec:perf-binning-method}). Note that the complexity of the binning method is only $O(M)$, where $M$ is the number of rows.

\subsection{Hashing method}\label{sec:moti-hashing-method}
The hashing method efficiently implements the accumulator in SpGEMM due to its good parallelism on GPUs~\cite{c_nsparse, c_speck}. Therefore, the hashing method is massively used in both $nsparse$ and $spECK$. Nevertheless, the hashing operation is one of the most time-consuming operations~\cite{c_nsparse} in SpGEMM.

One performance-critical operation in the hashing method is accessing the hash table with an inherently random access pattern. The inefficient implementation of the hashing method in $nsparse$ and $spECK$ is that they access the hash table too often, which may cause too many bank conflicts in the shared memory due to the random access. 

\subsection{Binning range selection}\label{sec:moti-binning-range}
The state-of-the-art SpGEMM libraries adopt row-wise and hashing methods to implement SpGEMM. One of the most important tasks in their implementation is choosing an appropriate hash table size to compute each output row. To make this choice appropriately, the trade-off between hash collision rate and hardware resource utilization needs to be carefully considered.
Specifically, the computing kernels are configured with a pre-defined hash table size. If the pre-defined hash table size in a kernel equals the largest $n_{prod}$ (or $n_{nz}$) of the rows computed by that kernel (potentially full occupancy of the hash table), the hash collision rate can be very high. However, the implementation can achieve high hardware resource utilization. In contrast, if the hash table size is kept unchanged, when computing rows with smaller $n_{prod}$ (or $n_{nz}$), it is observed that the hash collision rate decreases significantly. However, the hardware resource utilization also decreases.

Analytical models for choosing the appropriate hash table size to $n_{prod}$ (or $n_{nz}$) ratio could be difficult due to the entangled effects of hardware parameters and sparse matrix properties. However, to the best of our knowledge, the experimental exploration of this design choice was also missed in existing libraries. $nsparse$ computes the rows with potentially full occupancy of their hash table, yielding a huge hash collision rate. $spECK$ noticed the hash collision issue and used a larger hash table size for the numeric phase such that the largest occupancy of the hash table is 2/3. However, $spECK$ did not demonstrate if this configuration is experimentally best. This paper experimentally explores the design choice of the binning range selection and finds a general best configuration in section~\ref{sec:perf-binning-range}.

\subsection{Global memory usage for metadata}\label{sec:moti-minimize}
To perform SpGEMM efficiently on GPUs, recording some metadata in the global memory is necessary. For example, rows are classified into multiple bins/groups for good global load balance~\cite{c_nsparse, c_speck}. The classified row ids should be stored in the global memory as metadata to be used in the following computation steps.

Our micro-benchmarking shows that the global memory allocation is much more expensive than memory access in execution time. For example, the bandwidth of allocating 4MB of global memory is roughly 13.7GB/s, whereas accessing global memory with the same size achieves 124GB/s on the NVIDIA Tesla V100 GPUs. Therefore, minimizing the global memory usage should be considered and even prioritized when designing the SpGEMM algorithm. However, the existing SpGEMM libraries~\cite{c_nsparse, c_speck} allocated too much global memory for the metadata in their implementations, which affect their performance. 

In $nsparse$, the majority of metadata is used to store 1) the classified row ids in the binning method and 2) the $n_{prod}$ and $n_{nz}$ information. $nsparse$ allocates two arrays of size $M$ to store them. We observe that the array to store the $n_{prod}$ and $n_{nz}$ information can be reduced by sharing the $C.rpt$ array. In $spECK$, the majority of metadata is used to store the classified row ids in the binning method. Unlike $nsparse$, $spECK$ uses a two-dimensioned array of size $M*NUM\_BIN$ to store the classified row ids, where $NUM\_BIN$ is the number of the classified bins. $spECK$ allocates much more metadata than $nsparse$.

\subsection{Overlapping of global memory allocation and kernel execution}
\label{sec:moti-overlap}
By using NVIDIA’s profiling tool Nsight Systems~\cite{nsysdoc} in our GPU program profiling, we observe that when the host executes $cudaMalloc$~\cite{cudaAPIdoc} to allocate global memory, the already launched kernels can execute normally on the device without performance loss. In other words, the global memory allocation executed on the host machine can be parallelized (overlapped) with the kernel execution on the device machine. We denote this feature as the memory allocation and kernel execution parallelism. Note that executing $cudaMalloc$ on the host usually takes much longer time than launching a kernel.

Existing SpGEMM libraries~\cite{c_nsparse, c_speck}, however, did not utilize this parallelism feature to improve the performance of SpGEMM. For example, in the symbolic and numeric phases, both $nsparse$ and $spECK$ first allocate the global memory for hash tables and then launch other computing kernels. In typical benchmarks, the performance overhead of allocating the global memory is significant. Therefore, lack of memory allocation and kernel execution overlapping can waste a large amount of GPU resources.

\subsection{Load imbalance of the streaming multiprocessors}\label{sec:moti-SM}
NVIDIA's GPUs usually have a few dozens of streaming multiprocessors (SMs) that can accommodate hundreds to thousands of thread blocks simultaneously. Therefore, either the insufficient number of thread blocks or varying execution durations of the thread blocks in SpGEMM may cause the load imbalance of the SMs, which limits the performance of SpGEMM.

In the row-wise SpGEMM, each output row can be computed independently. Therefore, the execution order of the multiple kernels in the symbolic and numeric phases does not affect the accuracy of SpGEMM. $nsparse$ reported that they launched the kernels concurrently with multiple CUDA streams to tackle the possible insufficient number of thread blocks when computing small matrices~\cite{c_nsparse}. 

However, when $nsparse$ launched the kernel that computes the largest rows with global memory hash tables, it immediately called the $cudaFree$~\cite{cudaAPIdoc} to release the global memory before launching other kernels. Since $cudaFree$ implicitly invokes $cudaDeviceSynchronize$~\cite{cudaAPIdoc}, when the kernel that computes the largest rows is executing on a few SMs, many other kernels are not yet launched for execution, causing a significant load imbalance among the SMs. $spECK$ optimized this problem in their source code~\cite{speck_source} by calling the $cudaFree$ in the last phase. However, details of this optimization are missing in their paper~\cite{c_speck}.

\subsection{Full occupancy of the kernels}\label{sec:moti-occupancy}
When executing a GPU program, the high device memory access latency is hidden by switching warp execution within one streaming multiprocessor (SM). A kernel can have more opportunities to hide its memory access latency by having more warps located on one SM. The theoretical full occupancy of a kernel is achieved when the maximum number of threads are resident on one SM, which is 2048 in current NVIDIA GPUs~\cite{cudadoc}.

Although up to 2048 threads can be executed on one SM in theory, the number of computing resources (e.g., FP64 cores~\cite{volta}) is much less than 2048. Therefore, for a computation-bounded~\cite{roofline} kernel where few memory accesses are needed, a relatively low occupancy may achieve similar performance as compared to the full occupancy counterpart. However, since SpGEMM is a typical memory-bounded application with an irregular memory access pattern, occupancy is critical for its performance. 

Moreover, when profiling the occupancy of the existing SpGEMM libraries~\cite{c_nsparse, c_speck} by using NVIDIA’s profiling tool Nsight Compute~\cite{ncudoc}, we find that the achieved occupancy of the kernels in the symbolic and numeric phases is nearly always larger than 90\% when the kernel is configured with theoretical full occupancy. This practically shows that occupancy is playing a vital role in performing SpGEMM.

However, although $nsparse$ and $spECK$ mentioned the occupancy issue, they did not prioritize achieving theoretical full occupancy in the design goals of their kernels. As a result, many kernels in $nsparse$ and $spECK$ did not achieve the theoretical full occupancy~\cite{nsparse_source, speck_source}, limiting their performance.

\section{OpSparse}\label{sec:method}
To cope with the inefficient implementations of the existing SpGEMM libraries, we propose our optimized SpGEMM framework $OpSparse$ in this section. We first describe the overall computation flow of $OpSparse$ and then describe multiple optimizations.

\begin{figure}[h]
\centering
\includegraphics[width=0.45\textwidth]{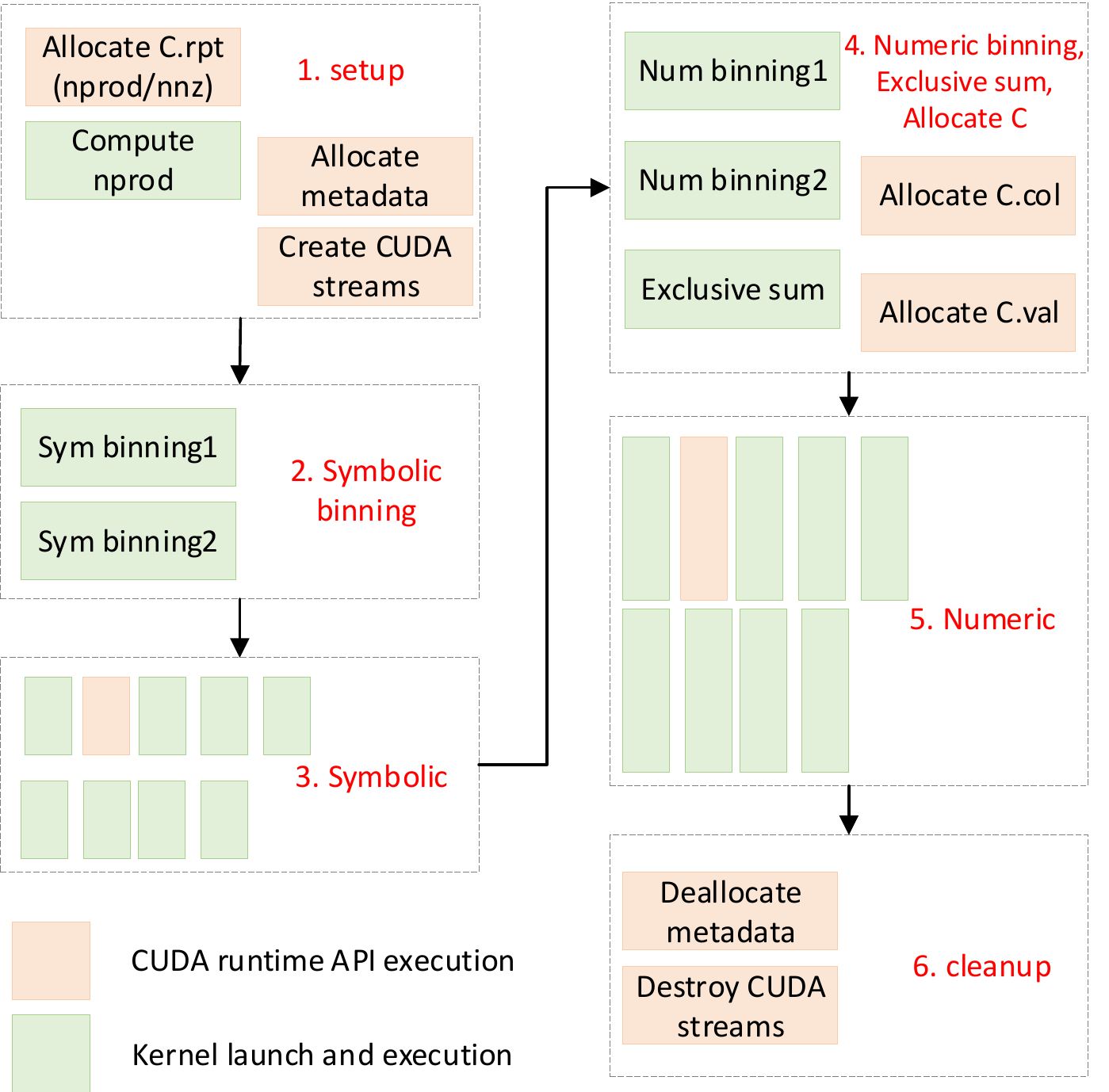}
\caption{Overall computation flow of $OpSparse$}
\label{fig:top}
\end{figure}

The overall computation flow is illustrated in Fig.~\ref{fig:top} with mainly six steps. Our computation flow is similar to $nsparse$ in that it adopts the row-wise, and two-phase methods to perform SpGEMM on GPUs. Two main computation steps are the symbolic and numeric steps (step3 and step5), which compute the $n_{nz}$, and the result column indices and values per output row, respectively. We only use hashing method to calculate the $n_{nz}$ and the result matrix in the two main computation steps. To achieve good global load balance, the two main computation steps are each equipped with a binning step (i.e., the symbolic binning step and the numeric binning step). Before the numeric step, we compute the total $n_{nz}$ and allocate the $C.col$ and $C.val$ arrays for the result matrix. We also compute the $C.rpt$ array by the exclusive-sum operation on the computed $n_{nz}$ information. In the setup step, we allocate the $C.rpt$ array, allocate the necessary device and host memory for the metadata used in SpGEMM, and compute the $n_{prod}$ per output row used in the symbolic binning step. Moreover, we create multiple CUDA streams~\cite{cudaAPIdoc} in the setup phase used for the concurrent kernel execution~\cite{cudadoc} in step3 and step5. In the final cleanup step, we release the allocated memory for the metadata and destroy the created CUDA streams.

Note that, in the symbolic and numeric steps, there are multiple kernels to compute rows with different $n_{prod}$ (or $n_{nz}$). We use multiple CUDA streams to launch these kernels concurrently.

\subsection{Efficiently utilizing the shared memory for the binning method}\label{sec:binning-method}
\begin{figure}[h]
\centering
\includegraphics[width=0.45\textwidth]{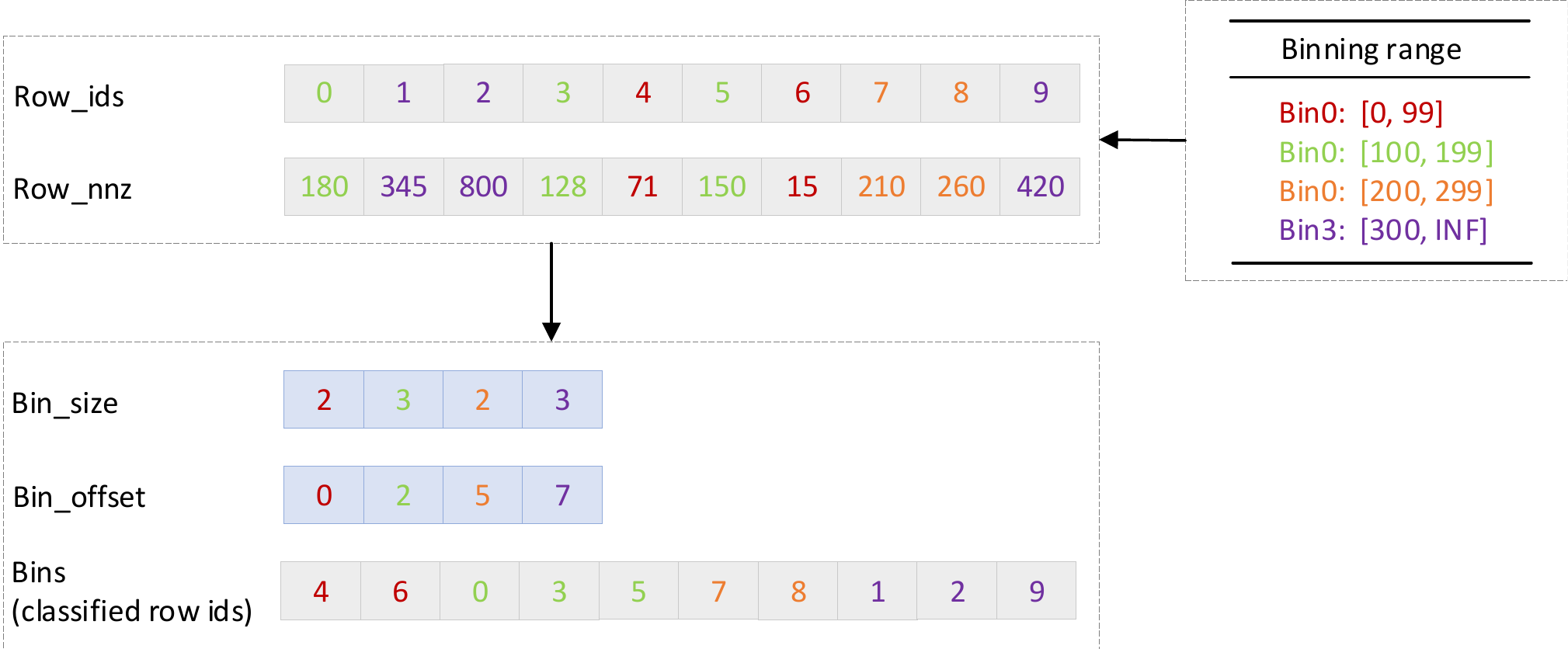}
\caption{Illustration of the binning method with minimized memory usage for the metadata. We use four bins and ten rows for the illustration. Top left (input data): ten rows with different $n_{nz}$ per row. Top right (input range): the binning ranges of the four bins. Bottom left (output data): The metadata to record the classified row ids ($bins$ array) and necessary indexing information. We use minimum entries to record all the classified row ids, which is ten. We also use the auxiliary $bin\_size$ and $bin\_offset$ arrays to record the number of rows, and their start offset in the $bins$ array.}
\label{fig:binning_method}
\end{figure}

The binning method is used in the two binning steps, which classify the rows into different bins for good global load balance. The binning method requires metadata to record its result, such as the classified row ids. Fig.~\ref{fig:binning_method} illustrates our implementation of the binning methods that minimizes the memory usage for the metadata.

Our implementations of the symbolic binning and numeric binning steps are the same, only with a difference in the input data. The symbolic binning step takes the $n_{prod}$ information as input, whereas the numeric binning step takes the $n_{nz}$ information. Therefore, the metadata used for the two binning steps is also shared to minimize the metadata usage further.

Since we aim to store the classified row ids in one array of length $M$ and we do not know each bin's size and offset (Fig.~\ref{fig:binning_method}), we need a two-pass computation method to implement the binning method. In the first pass (Algorithm~\ref{alg:binning1}), the bin sizes are counted. And then, the bin offsets are computed by exclusive-sum operation on the bin sizes. In the second pass (Algorithm~\ref{alg:binning2}), the offset position for each bin is known, and the classified row ids are written to their appropriate positions in the $bins$ array.

\begin{algorithm}
\caption{Binning method (first pass)}
\label{alg:binning1}
\begin{algorithmic}[1]
\State {// Set $d\_bin\_size$ to all zeroes before this kernel execution}
\State{$i\ =\ tid + blockIdx \cdot blockDim$}
\State {Initialize $r\_range[NUM\_BIN]$ in registers}
\State {Initialize $s\_bin\_size[NUM\_BIN]$ to zeros in shared memory}
\State {Initialize $s\_max\_row\_nnz[0]$ to zero}
\State{$atomicMax(s\_max\_row\_nnz,\ nnz[i])$}
\For {$j = 0\ to\ NUM\_BIN-1$}
    \If{$nnz[i] \leq r\_range[j]$}
        \State{$atomicAdd(s\_bin\_size + j,\ 1)$}
        \State{goto $Label$}
    \EndIf
\EndFor
\State{$Label$:}
\State{syncthreads}
\If{$tid < NUM\_BIN$}
    \State {$atomicAdd(d\_bin\_size + tid,\ s\_bin\_size[tid])$}
\EndIf
\If{$tid == 0$}
    \State {$atomicMax(d\_max\_row\_nnz,\ s\_max\_row\_nnz[0])$}
\EndIf
\end{algorithmic}
\end{algorithm}

\begin{algorithm}
\caption{Binning method (second pass)}
\label{alg:binning2}
\begin{algorithmic}[1]
\State {// Set $d\_bin\_size$ to all zeroes before this kernel execution}
\State {// $d\_bin\_offset$ is already computed}
\State {$i\ =\ tid + blockIdx * blockDim$}
\State {Initialize $r\_range[NUM\_BIN]$ in registers}
\State {Initialize $s\_bin\_size[NUM\_BIN]$ to zeros}
\For {$j = 0\ to\ NUM\_BIN-1$}
    \If{$nnz[i] \leq r\_range[j]$}
        \State{$atomicAdd(s\_bin\_size + j,\ 1)$}
        \State{goto $Label$}
    \EndIf
\EndFor
\State{$Label$:}
\State{syncthreads}
\State{// Compute s\_bin\_offset}
\If{$tid < NUM\_BIN$}
    \State{$s\_shared\_offset[tid]\ =\newline \ atomicAdd\ (d\_bin\_size + tid,\ s\_bin\_size[tid])$}
    \State{$s\_bin\_offset[tid]\ +=\ d\_bin\_offset[tid]$}
\EndIf
\State {Reset $s\_bin\_size[NUM\_BIN]$ to zeros}
\State{syncthreads}

\State{// Write row ids to $d\_bins$}
\For {$j = 0\ to\ NUM\_BIN-1$}
    \If{$nnz[i] \leq r\_range[j]$}
        \State{$r\_index = atomicAdd(s\_bin\_offset + j,\ 1)$}
        \State{$d\_bins[r\_index]\ =\ i$}
        \State{return}
    \EndIf
\EndFor

\end{algorithmic}
\end{algorithm}

Atomic operations should be used when computing the bin size and bin offset in the two binning kernels (Algorithm~\ref{alg:binning1} and Algorithm~\ref{alg:binning2}) since the bin size (Line 9 in Algorithm~\ref{alg:binning1}, Line 8 in Algorithm~\ref{alg:binning2}) and bin offset (Line 24 in Algorithm~\ref{alg:binning2}) are processed by multiple threads. Our algorithms fully utilize the shared memory to process the bin sizes and the bin offsets by atomic operations to improve the performance.

\begin{figure}
    \centering
    \includegraphics[width=0.45\textwidth]{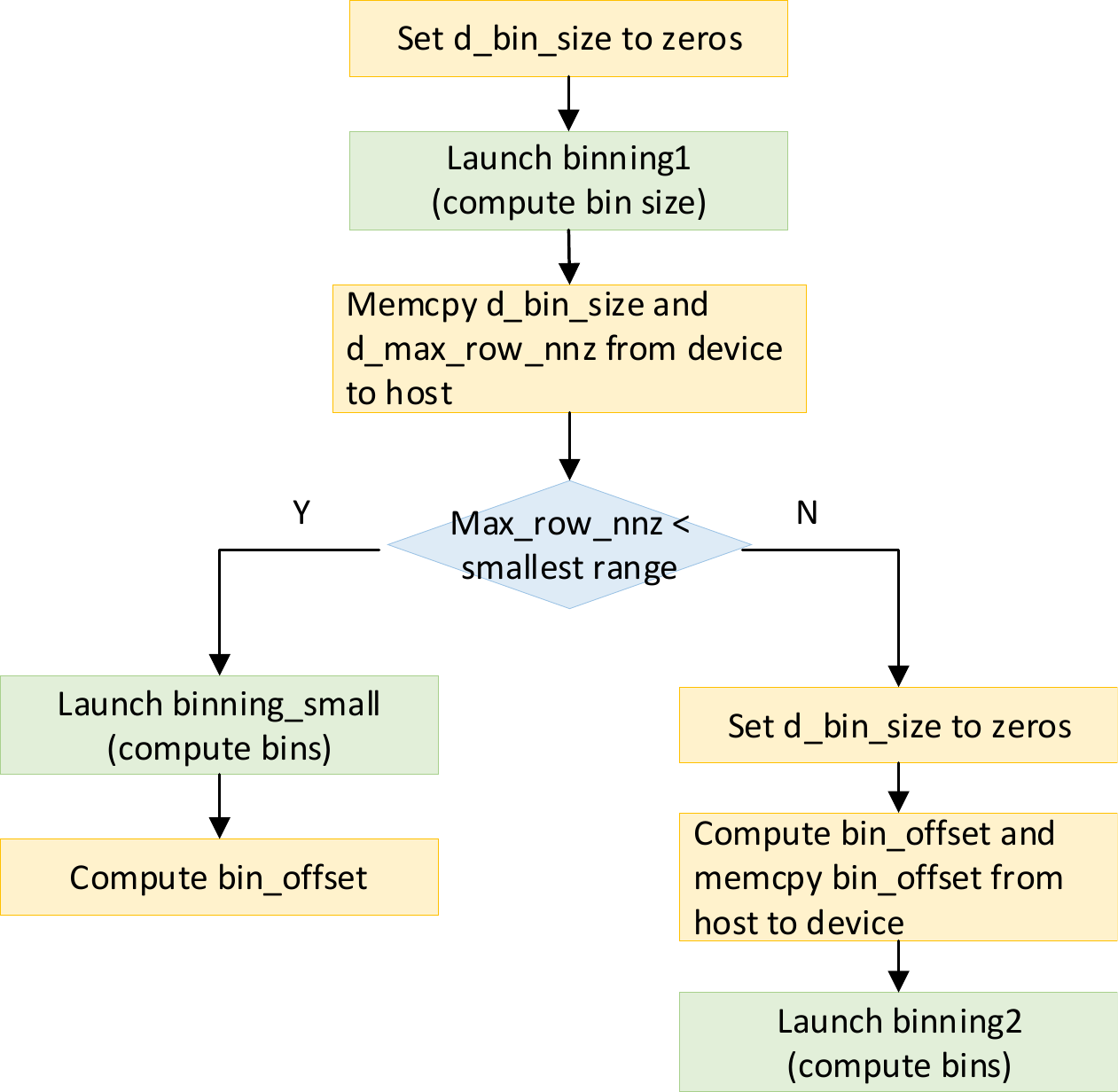}
    \caption{Overall computation flow of our binning method}
    \label{fig:binning_flow}
\end{figure}

Furthermore, we also utilize shared memory to monitor the max $n_{prod}$ (or $n_{nz}$) information in the first pass (Line 6 in Algorithm~\ref{alg:binning1}). If all the $n_{prod}$ (or $n_{nz}$) can be classified into the smallest bin (bin0), we save the comparison operations in the second pass by directly setting the $bins$ array as increasing row ids with another kernel (Algorithm~\ref{alg:binning_small}). The overall computation flow of our binning method is shown in Fig.~\ref{fig:binning_flow}.

\begin{algorithm}
\caption{Binning method (small)}
\label{alg:binning_small}
\begin{algorithmic}[1]
\State {$i\ =\ tid + blockIdx * blockDim$}
\State {$d\_bins[i]\ =\ i$}
\end{algorithmic}
\end{algorithm}

\subsection{Reducing the accesses to the hash table for the hashing method} \label{sec:hasing-method}
In the row-wise and hashing SpGEMM, each output row is computed with one hash table. The hash table is initialized to all $-1$ representing unoccupied status since the column indices ($key$) will be no less than $0$. To determine the address to insert the $key$ into the hash table, the $key$ is multiplied with a constant integer $HASH\_SCALE$, and then the result is mod-ed with the hash table size as the initial address ($hash$). However, the $hash$ of different $key$s can be the same, referred to as the hash collision issue. When the hash collision occurs, we increment the $hash$ by one until an unoccupied place in the hash table is found. Another situation is that there may be multiple duplicated $key$s when computing one result element. When duplicated $key$s are detected in the symbolic step, the $n_{nz}$ should only be incremented in the first occurrence of the duplicated $key$s. In contrast, when duplicated $key$s are detected in the numeric step, the values with the same duplicated $key$s should be accumulated together to generate one final element.

Algorithm~\ref{alg:hash_sym} and Algorithm~\ref{alg:hash_num} show our proposed hashing method in the symbolic and numeric steps, respectively. We introduce one optimization in our hashing method. In each iteration in the while loop, we access the hash table only once to minimize the access to shared memory hash tables, reducing the bank conflicts of the shared memory. We use atomic compare and swap ($atomicCAS$) to access the hash table and store the swapped result in a register, which is reused in the following computation.

\begin{algorithm}
\caption{Hashing method in the symbolic step}
\label{alg:hash_sym}
\begin{algorithmic}[1]
\State {Initialize $shared\_table$ to $-1$}
\State{Initialize $shared\_nnz[0]$ to $0$}
\State {Obtain the $key$ the same way as $nsparse$}
\State{$hash = key * HASH\_SCALE\ \&\ (t_{size} - 1)$}
\While {$true$}
    \State{$old = atomicCAS(shared\_table + hash, -1, key)$}
    \If{$old == -1$}
        \State{$atomicAdd(shared\_nnz, 1)$}
        \State{break}
    \ElsIf{$old == key$}
        \State{break}
    \Else
        \State{$hash = (hash + 1)\ \&\ (t_{size} - 1)$}
    \EndIf
\EndWhile
\end{algorithmic}
\end{algorithm}

\begin{algorithm}
\caption{Hashing method in the numeric step}
\label{alg:hash_num}
\begin{algorithmic}[1]
\State {Initialize $shared\_col$ to $-1$}
\State {Initialize $shared\_val$ to $0$}
\State {Obtain the $key$, $aval$, and $bval$ the same way as $nsparse$}
\State{$hash = key * HASH\_SCALE\ \%\ t_{size}$}
\While {$true$}
    \State{$old = atomicCAS(shared\_col + hash, -1, key)$}
    \If{$old == -1\ ||\ old == key$}
        \State{$atomicAdd(shared\_val + hash,\ aval * bval)$}
        \State{break}
    \Else
        \State{$hash = hash + 1 < t_{size}\ ?\ hash + 1\ :\ 0$}
    \EndIf
\EndWhile
\end{algorithmic}
\end{algorithm}

When the hash table size ($t_{size}$) in hashing method is a power-of-two number, the mod operation can be mathematically replaced with the logic-and operation~\cite{c_cusparse} (Line 4 and Line 13 in Algorithm~\ref{alg:hash_sym}). We utilize the logic-and operation since it is executed faster than the mod operation on GPUs. In the symbolic step, we keep most of the $t_{size}$s as power-of-two numbers and use the logic-and operation to improve performance. However, in the numeric step, we do not set the $t_{size}$ as power-of-two numbers for performance gains by full occupancy, detailed in section~\ref{sec:occupancy-num}. Consequently, we use mod operation in the numeric step (Line 4 in Algorithm~\ref{alg:hash_num}).

\subsection{Minimizing global memory usage for the metadata}\label{sec:minimize}
The metadata usage in our implementation only comes from the two binning steps and the exclusive-sum task in step5. How we minimize the metadata usage for the binning method has been described in section~\ref{sec:binning-method} when introducing the binning method. For the exclusive-sum task, we use the highly optimized $cub::DeviceScan::ExclusiveSum$ API~\cite{cubdoc} to implement it. The CUB library defines the required global memory usage to store its intermediate result. Our profiling shows that the size of the memory usage in the CUB library $cub::DeviceScan::ExclusiveSum$ can be computed with low cost ($\sim$1.2us) before computing the real results of the exclusive-sum operation~\cite{cubdoc}.

Although the computed $n_{prod}$ in the setup step and the computed $n_{nz}$ in the symbolic step also need memory to store their results, we save the memory usage of the $n_{prod}$ and $n_{nz}$ information by reusing the $C.rpt$ array. The reasons are four-fold. 1) The size of the $C.rpt$ array is easily obtained as $M + 1$, where $M$ is the number of rows of the $A$ matrix; therefore, the $C.rpt$ array can be allocated before computing the $n_{prod}$ and $n_{nz}$. 2) The $n_{prod}$ and $n_{nz}$ information are computed in different steps; therefore, they can be shared with the same memory space. 3) The memory size for $n_{prod}$ or $n_{nz}$ is $M$, which is smaller than the memory size of the $C.rpt$ array. 4) The $C.rpt$ array can be used by $n_{prod}$ or $n_{nz}$ since the $C.rpt$ array is not used for its original purpose until step4. Note that when computing the $C.rpt$ array in step4, the $cub::DeviceScan::ExclusiveSum$ library can compute the exclusive-sum operation in-place.


Furthermore, since the required size of all the metadata is known in our implementation before performing SpGEMM, we sum them up and allocate them together with one $cudaMalloc$ call to further reduce the overheads of global memory allocation. As a result, our memory usage for the metadata is self-managed.

\subsection{Overlapping memory allocation with kernel execution}\label{sec:overlap}
There are multiple computation steps where we can overlap the kernel execution with global memory allocation. In the setup step, we need to allocate global memory for metadata and compute the $n_{prod}$ information. The two tasks do not have a logical dependency. Therefore, we first launch the kernel to compute $n_{prod}$ and then allocate the global memory for the metadata without waiting for the kernel computation to finish. This execution sequence is shown in Fig.~\ref{fig:top}.

In step5, we first compute the total $n_{nz}$ within the first pass of the numeric binning method. Then we launch the second pass of the binning method and the exclusive sum interleaved with the two memory allocations for the $C.col$ and $C.val$ array (Fig.~\ref{fig:top}). Note that we make sure the exclusive-sum operation is computed after the two passes of the numeric binning step because we reuse the $C.rpt$ array to record the $n_{nz}$ information.

In the symbolic and numeric steps, there may be global memory allocation for the hash table to compute extremely large rows. Since the kernels that use shared memory hash tables do not depend on the global memory allocation, we first launch one kernel that use a shared memory hash table and then allocate the global memory immediately. As a result, part or all of the execution time of the memory allocation is parallelized with the pre-launched kernel execution.

\subsection{Load balance of the streaming multiprocessors}\label{sec:SM}
The load imbalance of SMs in our implementation may occur in the symbolic and numeric steps since rows are computed on different SMs with varying execution time.

One attribute of the concurrently-launched kernels is that the thread blocks in the earlier launched kernel still execute earlier than or concurrently with the thread blocks in
the later launched kernels. Therefore, to achieve a good load balance of the SMs, our implementation launches kernels that compute large rows relatively earlier and launch kernels with the smallest thread block size in the last phase. Moreover, we do not call $cudaFree$, which implicitly invokes $cudaDeviceSynchronize$ until all kernels have been launched with multiple CUDA streams.

\subsection{Kernel configuration}\label{sec:occupancy}
This section describes how we configure the kernel's parameters to achieve theoretical full occupancy for the symbolic and numeric steps. Nevertheless, some kernels are configured with theoretically 50\% occupancy for maximum shared memory usage.

Our implementation targets the NVIDIA Tesla V100  GPUs, which implement several dozens of SMs on one GPU. Each SM has maximumly 96KB shared memory, and the maximum thread block size for a kernel is 1024. For NVIDIA's GPU, the theoretical occupancy is affected by register usage, thread block size, and shared memory usage. We use the $\_\_launch\_bounds\_\_(1024, 2)$ compiler directive to force the register usage of the kernels to meet the requirement of theoretical full occupancy. We describe how we set each kernel's shared memory usage and thread block size as follows.

\subsubsection{Kernel configuration in the symbolic step}\label{sec:occupancy-sym}
We divide the rows with different $n_{prod}$ into 8 bins for computation by 9 kernels in the symbolic step. Table~\ref{tab:config_sym} shows the kernels' parameter settings and the $n_{prod}$ range in the symbolic step on NVIDIA Tesla V100 GPU. The rows in bin0--bin6 are computed by the corresponding kernel0--kernel6, and each bin is computed by one kernel. The rows in bin7 are firstly computed by kernel7. When the computed $n_{nz}$ of any row exceeds a threshold value, the row id is recorded, and the row is recomputed by kernel8. We set the threshold value as 0.8 times the kernel7's hash table size. kernel0--kernel7 implements shared memory hash tables, and kernel 8 implements global memory hash tables.

\begin{table}[h]
    \centering
    \caption{Parameter setting and binning ranges for the symbolic step on NVIDIA's Tesla V100 GPU}
    \label{tab:config_sym}
    \begin{tabular}{|c|c|c|c|c|}
        \hline
        \textbf{Bins} & \textbf{Kernels} & \textbf{Table size} &  \textbf{TB size} & \textbf{Ranges (Sym\_1.2x)}\\
        \hline
        Bin0 & Kernel0 & \textbf{32} & 4 $\cdot$ 256 = 1024 & 0 -- \textbf{26}  \\
        \hline
        Bin1 & Kernel1 & \textbf{512}  & 64  & 27 -- \textbf{426}\\
        \hline
        Bin2 & Kernel2 & \textbf{1024}  & 128  & 427 -- \textbf{853}\\
        \hline
        Bin3 & Kernel3 & \textbf{2048}  & 256  & 854 -- \textbf{1706}\\
        \hline
        Bin4 & Kernel4 & \textbf{4196}  & 512  & 1707 -- \textbf{3413}\\
        \hline
        Bin5 & Kernel5 & \textbf{8192}   & 1024 & 3414 -- \textbf{6826}\\
        \hline
        Bin6 & Kernel6 & \textbf{12287}  & 1024  & 6827 -- \textbf{10240}\\
        \hline
        Bin7 & \makecell{Kernel7 \\Kernel8} & \makecell{\textbf{24575}\\-}& \makecell{1024\\1024}  & \makecell{10241 -- $\infty$  \\ -}  \\
        \hline
    \end{tabular}
\end{table}

Kernel0 computes multiple rows in one thread block, which means there are multiple hash tables in one thread block. In contrast, kernel1--kernel8 compute one row in each thread block. We first describe the shared memory usage and thread block size configuration in kernel1. The kernel1's thread block size is the minimum thread block size, which is 64. Therefore, one SM can contain 2048/64 = 32 thread blocks, which requires 32 hash tables. Since each column index needs 4 bytes, the largest hash table size $t_{size}$ can be set as 96KB/32/4KB = 768. We set the $t_{size}$ as 512 to utilize the lightweight logic-and operation in the hashing method (Algorithm~\ref{alg:hash_sym}). We also use an additional 4 bytes of shared memory as $shared\_nnz$ to count the $n_{nz}$ per row. With the above configuration, kernel1 achieves theoretical full occupancy. Then we double the $t_{size}$ and thread block size in the following kernel2--kernel5 so that they all achieve theoretical full occupancy.

Kernel6 is configured with the maximum thread block size, which is 1024. It uses (48K-4) bytes and 4 bytes of shared memory to implement the hash table and $shared\_nnz$. Therefore, two thread blocks of 2048 threads can reside on one SM, achieving theoretical full occupancy.

Kernel7 uses the maximum shared memory (96KB) to implement the hash table and $shared\_nnz$, while the thread block size stays the maximum 1024. Therefore kernel7 achieves theoretically 50\% occupancy. Note that kernel7 computes the group of rows with the largest $n_{prod}$. If the actual $n_{nz}$ for a particular row exceeds a threshold size, that row will be recorded and recomputed in kernel8. We configure kernel7 with maximum hash table size to compute more large rows so that they are not computed by a much more expensive global memory hash table in kernel8.

Kernel8 only uses 4 bytes of shared memory to implement $shared\_nnz$, and the thread block size is set as 1024. Therefore, kernel8 achieves theoretical full occupancy.

Kernel0 computes multiple rows in one thread block. We use 4 threads to compute one row. Therefore, 2048/4 = 512 rows can be computed in one SM, which requires 512 hash tables on one SM. The largest $t_{size}$ is 96KB/512/4B = 48. Similar to the configuration in kernel1--kernel5, we set the $t_{size}$ as 32 and use an additional 4 bytes of shared memory to store the $shared\_nnz$ information.

\subsubsection{Kernel configuration in the numeric step}\label{sec:occupancy-num}
In the numeric step, we divide the rows with different $n_{nz}$ into 8 bins for computation by 8 kernels. Table~\ref{tab:config_num} shows the kernels' parameter settings and the $n_{nz}$ range on NVIDIA Tesla V100 GPU. The rows in bin0--bin7 are computed by the corresponding kernel0--kernel7, and each bin is computed by one kernel. kernel0--kernel6 implements shared memory hash tables, and kernel7 implements global memory hash tables.

\begin{table}[h]
    \centering
    \caption{Parameter setting and binning ranges for the numeric step on NVIDIA's Tesla V100 GPU}
    \label{tab:config_num}
    \begin{tabular}{|c|c|c|c|c|}
        \hline
        \textbf{Bins} & \textbf{Kernels} & \textbf{Table size} &  \textbf{TB size} & \textbf{Ranges (Num\_2x)}\\
        \hline
        Bin0 & Kernel0 & \textbf{31}  & 8 $\cdot$ 128 = 1024  & 0 -- \textbf{16}\\
        \hline
        Bin1 & Kernel1 & \textbf{255}  & 64  & 17 -- \textbf{128} \\
        \hline
        Bin2 & Kernel2 & \textbf{511}  & 128  & 129 -- \textbf{256} \\
        \hline
        Bin3 & Kernel3 & \textbf{1023}  & 256  & 257 -- \textbf{512}\\
        \hline
        Bin4 & Kernel4 & \textbf{2047}  & 512  & 513 -- \textbf{1024}\\
        \hline
        Bin5 & Kernel5 & \textbf{4095}  & 1024  & 1025 -- \textbf{2048}\\
        \hline
        Bin6 & Kernel6 & \textbf{8191} & 1024  & 2049 -- \textbf{4096} \\
        \hline
        Bin7 & Kernel7 & -  & 1024 & 4097 -- $\infty$ \\
        \hline
    \end{tabular}
\end{table}

The kernels in the numeric phase consist of three computing phases: hashing, condensing, and sorting. The hashing phase computes the values and column indices of the output matrix by utilizing the hash table. In double-precision, each value and column index pair requires 12 bytes of memory. After the hashing phase, the values and column indices are stored sparsely in the hash table being unsorted. The condensing phase gathers the sparsely stored values and column indices with multiple threads. When a thread carries valid data, it atomically increments an offset variable and then stores the valid data to the position pointed by the offset variable. We use 4 bytes of shared memory to implement this offset variable named $shared\_offset$. The last sorting phase sorts the nonzero values in the condensed hash map and stores them in global memory.

Kernel0 computes multiple rows in one thread block, which means it implements multiple hash tables in one thread block. In contrast, kernel1--kernel7 computes one row in each thread block. We first describe the shared memory usage and thread block size configuration in kernel1. The kernel1's thread block size is set as the minimum thread block size, which is 64. Therefore, one SM can contain 2048/64 = 32 thread blocks, which requires 32 hash tables. Since each value and column index pair requires 12 bytes in double precision. The largest table size $t_{size}$ is 96KB/64/12B = 256. Our algorithm sets the $t_{size}$ in kernel1 as 255, so that additional 4 bytes of shared memory can be used to implement the $shared\_offset$. With the above configuration, kernel1 achieves theoretical full occupancy. The kernel2--kernel5 are configured similarly to kernel1 with doubled $t_{size}$ and thread block size. Therefore, kernel2--kernel5 all achieve theoretical full occupancy.

Kernel6 uses the maximum shared memory (96K bytes) to implement the hash table and $shared\_offset$, while the thread block size is the maximum 1024. Therefore kernel6 achieves theoretical 50\% occupancy. We sacrifice half occupancy in kernel6 so that more rows can be computed by kernel6 with the shared memory hash table rather than kernel7 with the global memory hash table.

Kernel7 only uses 4 bytes of shared memory to implement $shared\_offset$, and the thread block size is set as 1024. Therefore, Kernel7 achieves theoretical full occupancy.

Kernel0 computes multiple rows in one thread block. Our algorithm sets 8 threads to compute each row. Therefore, maximum 2048/8 = 256 rows are computed in one SM, which requires 256 hash tables. The largest $t_{size}$ is 96KB/256/12B = 32. Similar to kernel1-kernel5, we set the $t_{size}$ as 31 and use the additional 4 bytes of shared memory to implement the $shared\_offset$ information. Kernel0 achieves theoretical full occupancy.

\subsection{Binning range selection}\label{sec:binning-range}
We have implemented the binning method and determined the hash table size and thread block size for all the computation kernels. This section describes the binning range selection with which each row with varying $n_{prod}$ (or $n_{nz}$) is assigned to the appropriate kernel. 

In the symbolic step, we keep the hash table size of each kernel at least $1.2\times$ larger than the largest $n_{prod}$ in the corresponding group. In the numeric step, we keep the hash table size of each kernel at least $2\times$ larger than the largest $n_{nz}$ in the corresponding group. Our binning ranges for symbolic and numeric steps are given in Table~\ref{tab:config_sym} and Table~\ref{tab:config_num}, respectively. These binning ranges are demonstrated with generally better performance than other binning ranges through our experiments which will be shown in section~\ref{sec:perf-binning-range}.

\section{Performance evaluation}\label{sec:evaluate}
We compare the overall performance with three state-of-the-art SpGEMM libraries $cuSPARSE$, $nsparse$, and $spECK$. We also provide individual performance analysis of our multiple optimizations and mainly compare them with ourselves in different implementation versions. The evaluation is based on the FLOPS performance of the matrix square benchmarks~\cite{c_bhsparse, c_nsparse, c_speck}, which is twice the number of the intermediate products divided by the execution time. The execution time is obtained by first executing the program once for warmup, then executing the program ten times to obtain an average execution time. In all benchmarks, we measure the execution time of the approaches in double precision.

\subsection{Evaluation environments}\label{sec:environment}
We select 26 square matrices from the SuiteSparse Matrix Collection~\cite{suitesparse}, which are commonly used for the performance evaluation of SpGEMM on GPUs~\cite{c_bhsparse, c_nsparse, c_speck}. The detailed information of these matrices are summarized in Table~\ref{tab:matrix}. Due to limited device memory capacity, $cuSPARSE$ cannot compute 7 matrices at the bottom in Table~\ref{tab:matrix}. Therefore, we denote the top 19 matrices in Table~\ref{tab:matrix} as normal matrices and the bottom 7 matrices in Table~\ref{tab:matrix} as large matrices.

\begin{table*}[h]
    \centering
    \fontsize{7.5}{12}\selectfont
    \caption{Detailed information of the 26 matrices. The top 19 matrices are denoted as normal matrices, whereas the bottom 7 matrices ars denoted as large matrices.}
    \label{tab:matrix}
    \begin{tabular}{|c|c|c|c|c|c|c|c|c|}
    \hline
    \textbf{Id} & \textbf{Name} & \textbf{Rows} & \textbf{Nnz} & \textbf{Nnz/row} & \textbf{Max nnz/row} & \textbf{Nprod of $A^2$} & \textbf{Nnz of $A^2$} & \textbf{Compression ratio of $A^2$} \\
    \hline
    1 & m133-b3 & \num{200200} & \num{800800} & 4.0 & \num{4} & \num{3203200} & \num{3182751} & 1.01 \\
    \hline
    2 & mac\_econ\_fwd500 & \num{206500} & \num{1273389} & 6.2 & \num{44} & \num{7556897} & \num{6704899} & 1.13 \\
    \hline
    3 & patents\_main & \num{240547} & \num{560943} & 2.3 & \num{206} & \num{2604790} & \num{2281308} & 1.14 \\
    \hline
    4 & webbase-1M & \num{1000005} & \num{3105536} & 3.1 & \num{4700} & \num{69524195} & \num{51111996} & 1.36 \\
    \hline
    5 & mc2depi & \num{525825} & \num{2100225} & 4.0 & \num{4} & \num{8391680} & \num{5245952} & 1.60 \\
    \hline
    6 & scircuit & \num{170998} & \num{958936} & 5.6 & \num{353} & \num{8676313} & \num{5222525} & 1.66 \\
    \hline
    7 & mario002 & \num{389874} & \num{2101242} & 5.4 & \num{7} & \num{12829364} & \num{6449598} & 1.99 \\
    \hline
    8 & cage12 & \num{130228} & \num{2032536} & 15.6 & \num{33} & \num{34610826} & \num{15231874} & 2.27 \\
    \hline
    9 & majorbasis & \num{160000} & \num{1750416} & 10.9 & \num{11} & \num{19178064} & \num{8243392} & 2.33 \\
    \hline
    10 & offshore & \num{259789} & \num{4242673} & 16.3 & \num{31} & \num{71342515} & \num{23356245} & 3.05 \\
    \hline
    11 & 2cubes\_sphere & \num{101492} & \num{1647264} & 16.2 & \num{31} & \num{27450606} & \num{8974526} & 3.06 \\
    \hline
    12 & poisson3Da & \num{13514} & \num{352762} & 26.1 & \num{110} & \num{11768678} & \num{2957530} & 3.98 \\
    \hline
    13 & filter3D & \num{106437} & \num{2707179} & 25.4 & \num{112} & \num{85957185} & \num{20161619} & 4.26 \\
    \hline
    14 & mono\_500Hz & \num{169410} & \num{5036288} & 29.7 & \num{719} & \num{204030968} & \num{41377964} & 4.93 \\
    \hline
    15 & conf5\_4-8x8-05 & \num{49152} & \num{1916928} & 39.0 & \num{39} & \num{74760192} & \num{10911744} & 6.85 \\
    \hline
    16 & cant & \num{62451} & \num{4007383} & 64.2 & \num{78} & \num{269486473} & \num{17440029} & 15.45 \\
    \hline
    17 & consph & \num{83334} & \num{6010480} & 72.1 & \num{81} & \num{463845030} & \num{26539736} & 17.48 \\
    \hline
    18 & shipsec1 & \num{140874} & \num{7813404} & 55.5 & \num{102} & \num{450639288} & \num{24086412} & 18.71 \\
    \hline
    19 & rma10 & \num{46835} & \num{2374001} & 50.7 & \num{145} & \num{156480259} & \num{7900917} & 19.81 \\
    \hline
    \hline
    20 & delaunay\_n24 & \num{16777216} & \num{100663202} & 6.0 & \num{26} & \num{633914372} & \num{347322258} & 1.83 \\
    \hline
    21 & cage15 & \num{5154859} & \num{99199551} & 19.2 & \num{47} & \num{2078631615} & \num{929023247} & 2.24 \\
    \hline
    22 & wb-edu & \num{9845725} & \num{57156537} & 5.8 & \num{3841} & \num{1559579990} & \num{630077764} & 2.48 \\
    \hline
    23 & cop20k\_A & \num{121192} & \num{2624331} & 21.7 & \num{81} & \num{79883385} & \num{18705069} & 4.27 \\
    \hline
    24 & hood & \num{220542} & \num{10768436} & 48.8 & \num{77} & \num{562028138} & \num{34242180} & 16.41 \\
    \hline
    25 & pwtk & \num{217918} & \num{11634424} & 53.4 & \num{180} & \num{626054402} & \num{32772236} & 19.10 \\
    \hline
    26 & pdb1HYS & \num{36417} & \num{4344765} & 119.3 & \num{204} & \num{555322659} & \num{19594581} & 28.34 \\
    \hline

    \end{tabular}
\end{table*}

For a fair comparison, we evaluate the performance of both our approach and others with the same hardware and software settings. NVIDIA Tesla V100 PCI-e GPU is used for the evaluation, which has 16GB device memory, and up to 900GB/s memory bandwidth. The code is compiled by nvcc with version 11.2, and the optimization level is O3. The code runs on the ubuntu 18.04.5 LTS operating system on a Intel Xeon Platinum 8163 CPU.

\subsection{Performance of SpGEMM}\label{sec:perf-SpGEMM}
Since $spECK$ does not include the overheads of allocating the $C$ matrix in their benchmarking~\cite{c_speck}, we slightly modify their source code to include the overheads of allocating $C$. Note that the overheads of allocating $C$ is included in all the other SpGEMM libraries, including ours. The $nsparse$'s source code~\cite{nsparse_source} provides a Volta version of their SpGEMM implementation. We use their Volta version for the performance evaluation. The source code of our SpGEMM library and the SpGEMM libraries of $spECK$, $nsparse$, and $cuSPARSE$ will be provided in a github repo after the acceptance by IEEE Access.

\begin{figure}[h]
\centering
\includegraphics[width=0.48\textwidth]{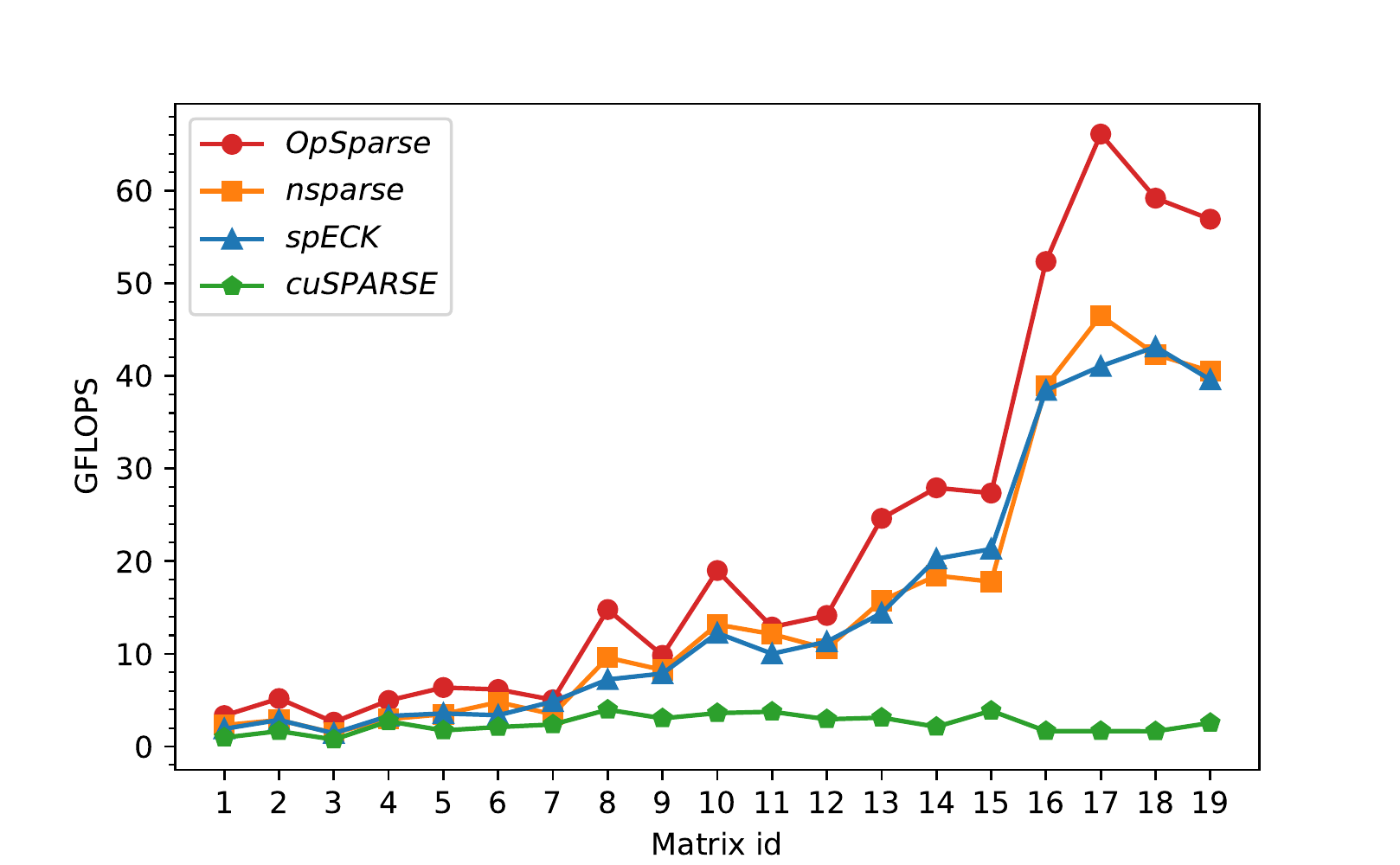}
\caption{Performance comparison of SpGEMM on 19 normal matrices}
\label{fig:normal}
\includegraphics[width=0.48\textwidth]{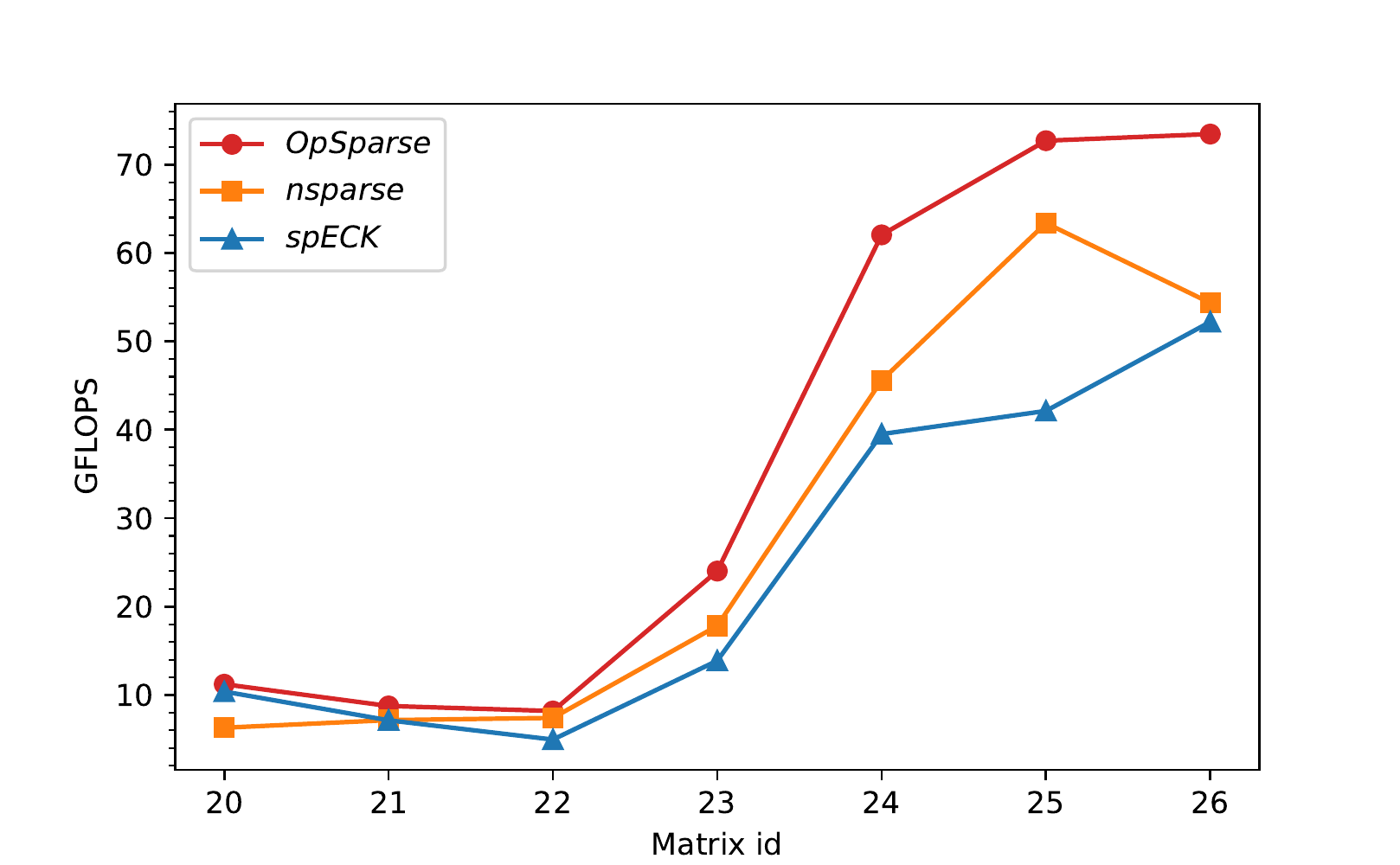}
\caption{Performance comparison of SpGEMM on 7 large matrices}
\label{fig:large}
\end{figure}

Fig.~\ref{fig:normal} shows the overall performance of SpGEMM for the normal matrices. We can see that $OpSparse$ clearly outperforms other state-of-the-art SpGEMM libraries on all the normal matrices. Specifically, $OpSparse$ performs on average $7.35\times$, $1.43\times$, and $1.52\times$ better than $cuSPARSE$, $nsparse$, and $spECK$, respectively. $cuSPARSE$ achieves relatively stable performance with varying input matrices. However, $cuSPARSE$'s performance is the lowest. $nsparse$ and $spECK$ achieve similar performance on all the normal matrices.

Fig.~\ref{fig:large} shows the overall performance of SpGEMM for the 7 large matrices. The performance of $cuSPARSE$ is omitted since it cannot compute the large matrices. $OpSparse$'s performance is the best among other libraries on all the 7 large matrices.


\subsection{Performance analysis of the multiple optimizations}\label{sec:perf-optimizations}
Our proposed framework provides multiple optimizations for SpGEMM. To better show the performance contribution of the optimizations, we analyze their performance improvements individually in this section.

\subsubsection{Binning method}\label{sec:perf-binning-method}
\begin{figure}[h]
\centering
\includegraphics[width=0.48\textwidth]{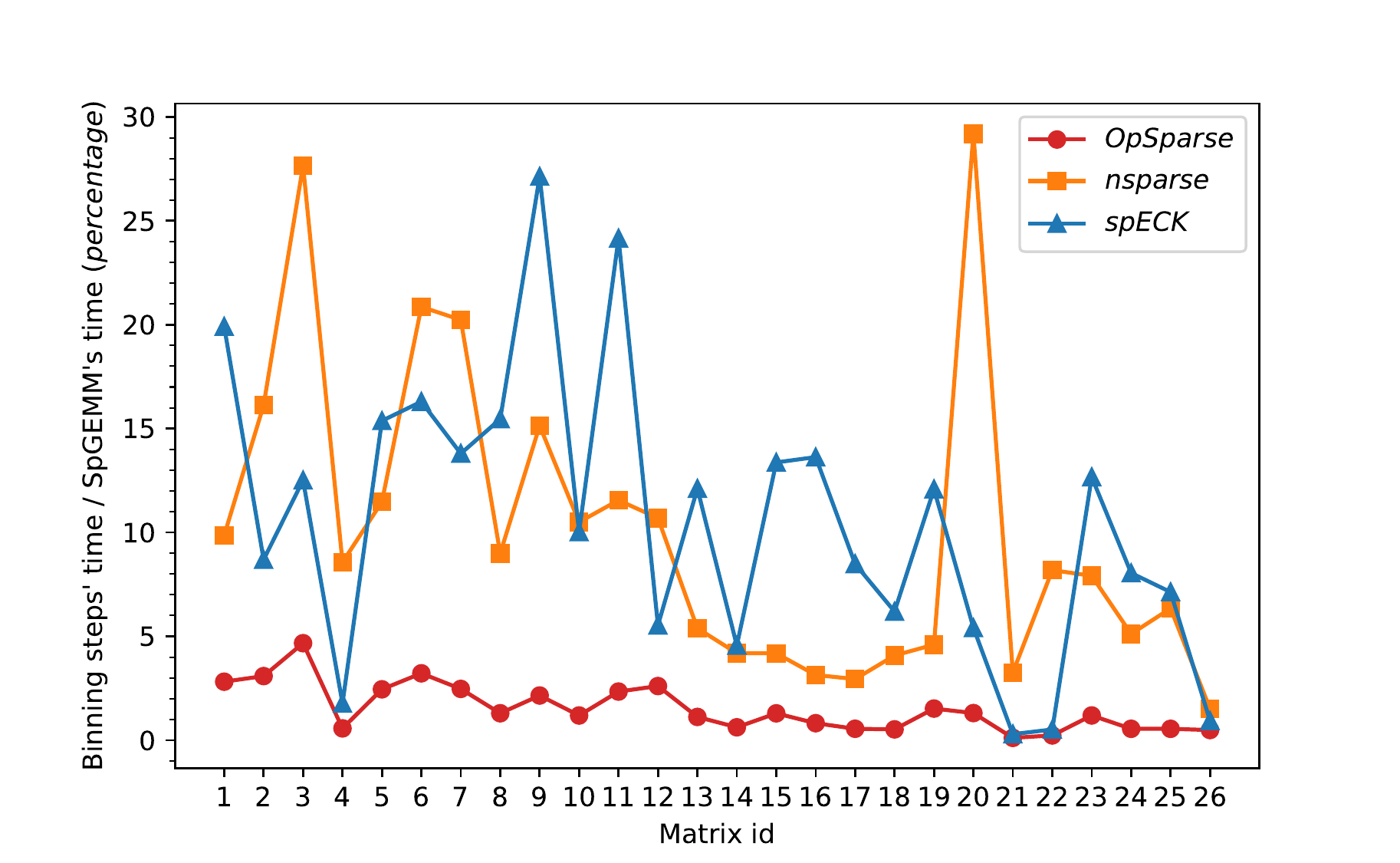}
\caption{Execution time of the two binning steps compared to the overall execution time of SpGEMM}
\label{fig:binning_percent}

\includegraphics[width=0.48\textwidth]{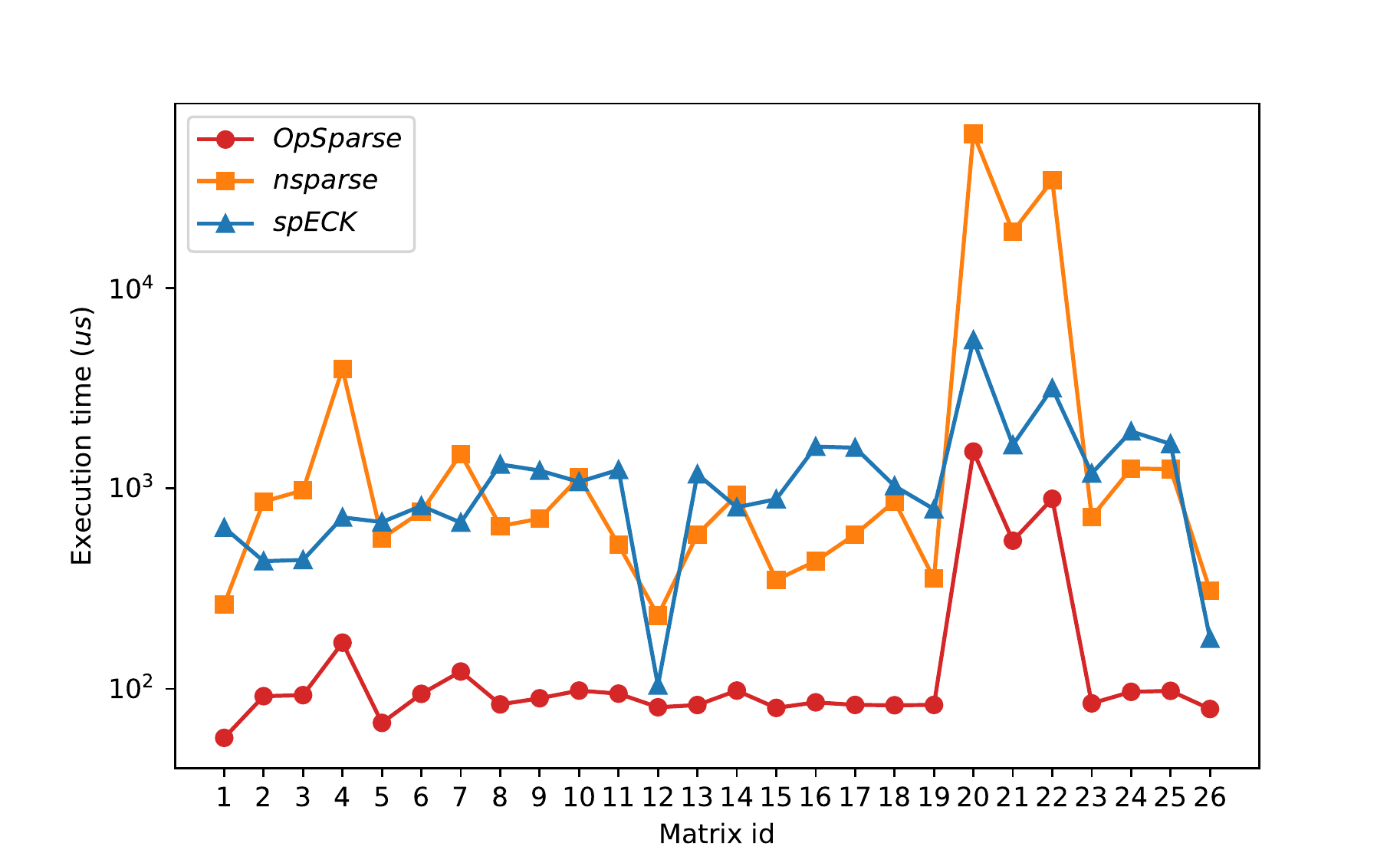}
\caption{Execution time of the two binning steps}
\label{fig:binning_time}
\end{figure}

We compare the performance of the binning method with $nsparse$ and $spECK$. Fig.~\ref{fig:binning_percent} shows the execution time of the two binning steps compared to the overall execution time of SpGEMM. $OpSparse$ clearly outperforms $nsparse$ and $spECK$. For $nsparse$, the two binning steps take up to 29.2\% of the total execution time in the worst case and 10.1\% of the total execution time on average for the 26 matrices. For $spECK$, the two binning steps take up to 27.1\% of the total execution time in the worst case and 10.6\% of the total execution time on average for the 26 matrices. In $OpSparse$, the two binning methods only take 4.7\% of the total execution time in the worst case and 1.5\% on average. The overhead of the two binning steps in $nsparse$ and $spECK$ is nontrivial since the binning steps are only a low complexity and auxiliary task for SpGEMM. By fully utilizing shared memory, our binning methods only take a little computation time in computing the overall SpGEMM.


Figure~\ref{fig:binning_time} shows the execution time of the two binning steps. The performance of the two binning steps in $OpSparse$ achieves on average $12\times$ and $10\times$ performance speedup compared to $nsparse$ and $spECK$, respectively. We attribute this impressive performance speedup to our better utilization of the shared memory.


\subsubsection{Hashing method}\label{sec:perf-hashing-method}
\begin{figure}[h]
\centering
\includegraphics[width=0.48\textwidth]{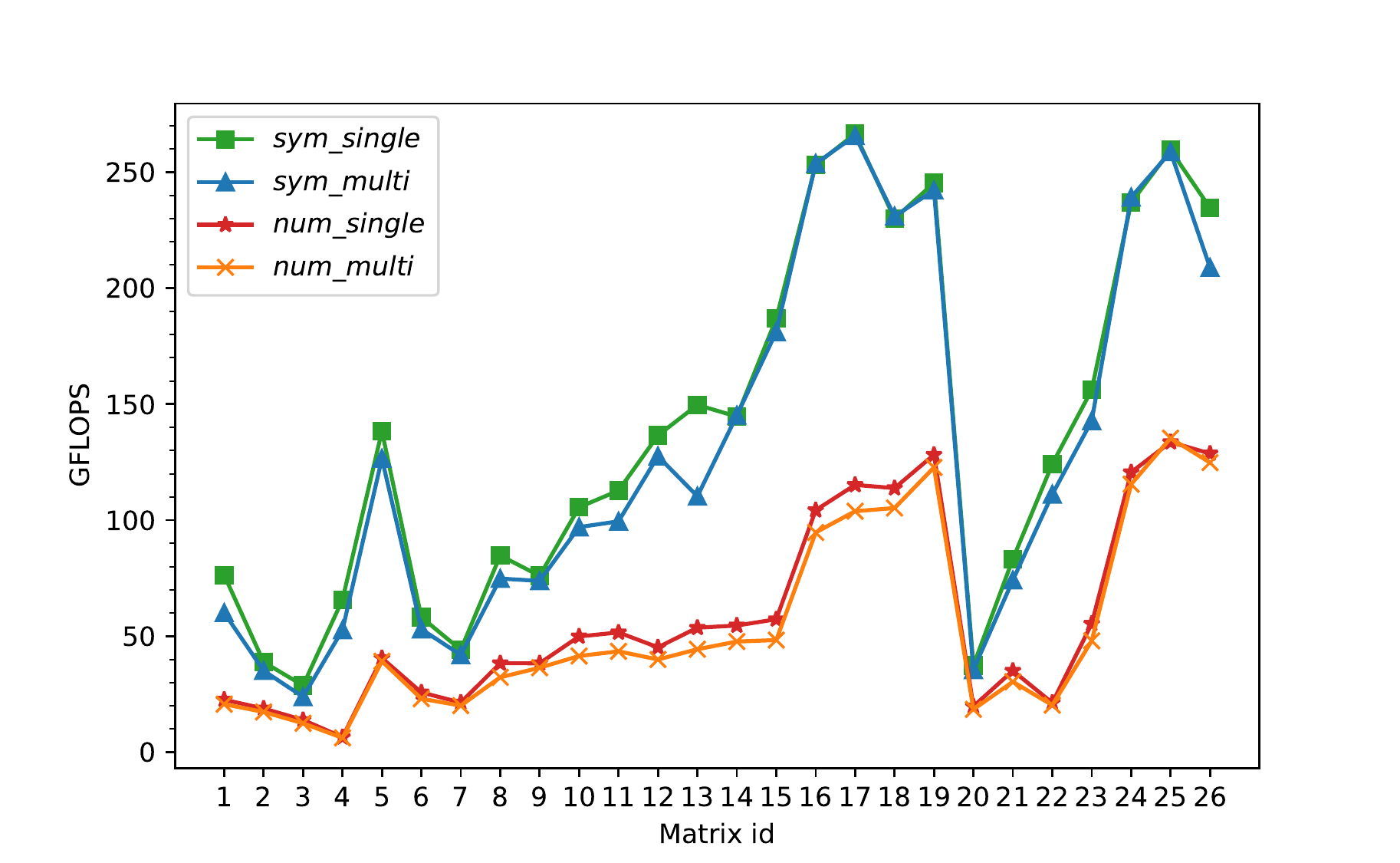}
\caption{Performance of the the symbolic and numeric steps with the two versions of the hashing method. The prefix $sym$ and $num$ denote the symbolic and numeric steps. The suffix $single$ and $multi$ denote the single access and multiple access versions of the hashing method.}
\label{fig:hash}
\end{figure}

Our hashing method is shown in Algorithm~\ref{alg:hash_sym} and Algorithm~\ref{alg:hash_num}. We optimise the access pattern to the hash tables compared to the traditional implementation in $nsparse$ and $spECK$. We access the hash table once in each while loop, whereas $nsparse$ and $spECK$ access the hash table multiple times in each while loop. To compare the performance difference between the two hashing versions, we implement the two hashing versions in our framework such that all other implementations are kept the same except for the access pattern. Fig.~\ref{fig:hash} shows the performance of the two versions of hashing methods in symbolic and numeric steps. Compared to the multiple access version of the hashing method, the single access version achieves on average $1.09\times$ and $1.10\times$ performance speedup in the symbolic step and numeric step, respectively. In $OpSparse$, we adopt the single access version of hashing method.

\subsubsection{Binning range selection}\label{sec:perf-binning-range}
Each kernel's hash table size is fixed in the symbolic and numeric steps in $OpSparse$. However, assigning rows with different $n_{prod}$ (or $n_{nz}$) to the kernels is a relatively free choice controlled by the binning ranges. As mentioned in section~\ref{sec:moti-binning-range}, selecting different binning ranges is a trade-off between hardware utilization efficiency and hash collision rate.

To explore which kinds of binning ranges yield the best performance for the symbolic and numeric steps, we test three kinds of binning ranges (shown in Table~\ref{tab:sym_range}) in the symbolic step and four kinds of binning ranges (shown in Table~\ref{tab:num_range}) in the numeric step. 

\begin{table*}[h]
    \centering
    \caption{The hash table size and three kinds of binning ranges for the symbolic step. The three binning ranges are obtained by keeping the hash table size of each kernel $1\times$, $1.2\times$, and $1.5\times$ larger than the largest $n_{prod}$ in the corresponding group. We denote the three binning ranges for the symbolic step as $sym\_1\times$, $sym\_1.2\times$, and $sym\_1.5\times$, respectively.}
    \label{tab:sym_range}
    \begin{tabular}{|c|c|c|c|c|}
        \hline
        \textbf{Kernels} & \textbf{Table size} & \textbf{Sym\_1x} & \textbf{Sym\_1.2x} & \textbf{Sym\_1.5x}\\
        \hline
        Kernel0 & \textbf{32} & 0 -- 32 & \textbf{0 -- 26} & 0 -- 21 \\
        \hline
        Kernel1 & \textbf{512} & 33-- 512  & \textbf{27 -- 426} & 22--341\\
        \hline
        Kernel2 & \textbf{1024} & 513 --1024 & \textbf{427 -- 853} & 341 -- 682 \\
        \hline
        Kernel3 & \textbf{2048} & 1025-- 2048 & \textbf{854 -- 1706} & 683-- 1365\\
        \hline
        Kernel4 & \textbf{4096} & 2049 --4096 & \textbf{1707 -- 3413} & 1366 -- 2730\\
        \hline
        Kernel5 & \textbf{8192} & 4197 -- 8192  & \textbf{3414 -- 6826} & 2731 -- 5461\\
        \hline
        Kernel6 & \textbf{12287} & 8193 -- 12287 & \textbf{6827 -- 10240} & 5462 -- 8191\\
        \hline
        Kernel7 & \textbf{24575} & 12288 -- $\infty$  & \textbf{10241 -- $\infty$} & 8192 -- $\infty$\\
        \hline
        
    \end{tabular}
    
\end{table*}

\begin{table*}[ht]
    \centering
    \caption{The hash table size and four kinds of binning ranges for the numeric computation. The four binning ranges are obtained by keeping the hash table size of each kernel $1\times$, $1.5\times$, $2\times$, and $3\times$ larger than the largest $n_{nz}$ in the corresponding group. We denote the four binning ranges for the numeric step as $num\_1\times$, $num\_1.5\times$, $num\_2\times$, and $num\_3\times$, respectively.}
    \label{tab:num_range}
    \begin{tabular}{|c|c|c|c|c|c|}
        \hline
        \textbf{Kernels} & \textbf{Table size} & \textbf{Num\_1x} & \textbf{Num\_1.5x} & \textbf{Num\_2x} & \textbf{Num\_3x}\\
        \hline
        Kernel0 & \textbf{31} & 0 -- 31 & 0 -- 21 & \textbf{0 -- 16} & 0 -- 10\\
        \hline
        Kernel1 & \textbf{255} & 32-- 255  & 22 -- 192 & \textbf{17--128} & 11--85\\
        \hline
        Kernel2 & \textbf{511} & 256 --511 & 193 -- 384 & \textbf{129 -- 256} & 86 -- 170\\
        \hline
        Kernel3 & \textbf{1023} & 512-- 1023 & 385 -- 768 & \textbf{257-- 512} & 171-- 341\\
        \hline
        Kernel4 & \textbf{2047} & 1024 --2047 & 769 -- 1536 & \textbf{513 -- 1024} & 342 -- 682\\
        \hline
        Kernel5 & \textbf{4095} & 2048 -- 4095  & 1537 -- 3072 & \textbf{1025 -- 2048} & 683 -- 1365\\
        \hline
        Kernel6 & \textbf{8191} & 4196 -- 8191 & 3073 -- 5460 & \textbf{2049 -- 4096} & 1366 -- 2730\\
        \hline
        Kernel7 & \textbf{-} & \textbf{8192 -- $\infty$}  & 5461 -- $\infty$ & \textbf{4097 -- $\infty$} & 2731 -- $\infty$\\
        \hline
        
    \end{tabular}
    
\end{table*}

\begin{figure}[h]
\centering
\includegraphics[width=0.48\textwidth]{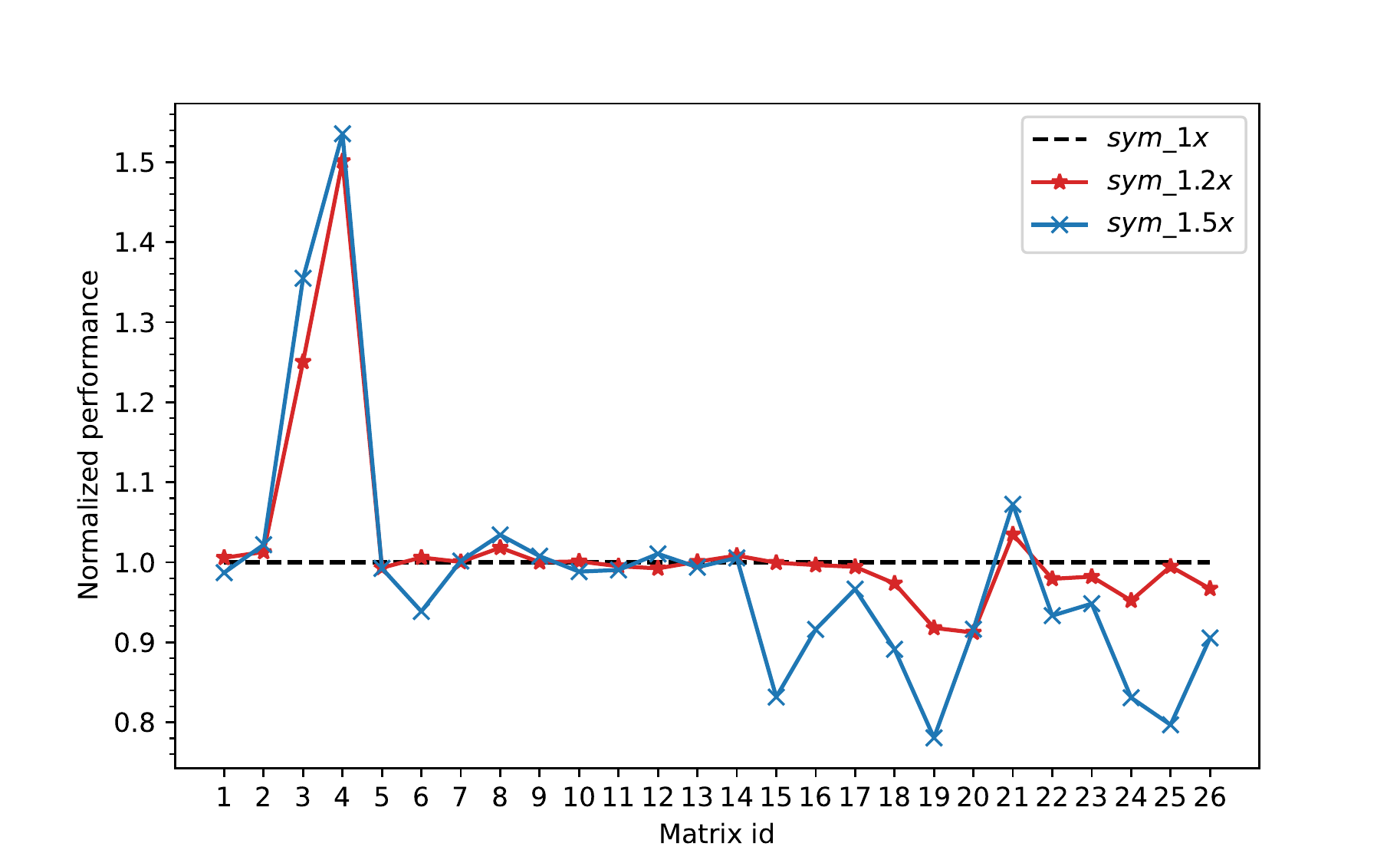}
\caption{Normalized performance of the symbolic step with different binning ranges. The performance is normalized to the performance with $sym\_1\times$ binning range. $sym\_1\times$, $sym\_1.2\times$, and $sym\_1.5\times$ denote the performance of the symbolic step with the $sym\_1\times$, $sym\_1.2\times$, and $sym\_1.5\times$ binning ranges (shown in Table~\ref{tab:sym_range}), respectively.}
\label{fig:sym_range}
\end{figure}

Fig.~\ref{fig:sym_range} shows the performance of the symbolic step with the three kinds of binning ranges. The implementations with the $sym\_1.2\times$ and $sym\_1.5\times$ binning ranges achieve on average $1.02\times$ and $0.99\times$ performance speedup, respectively, compared to the $sym\_1\times$ binning range. This means the average performance improvement when scaling down the binning range in the symbolic step is not significant and even negative. The reason is that, by using the $n_{prod}$ in the classification, the hash collision rate is usually low since the $n_{prod}$ is usually larger than the actual $n_{nz}$. 

However, scaling down the binning ranges may improve the performance nontrivially for benchmarks with a small compression ratio. For the matrix \emph{webbase-1M} and \emph{patents\_main}, the performance speedup of $sym\_1.2\times$ and $sym\_1.5\times$ binning range is significant compared to the $sym\_1\times$ binning range. We attribute the significant performance improvement to the lowered hash collision rate since the compression ratios of the \emph{webbase-1M} and \emph{patents\_main} benchmarks are only 1.36 and 1.15, respectively. For other tested benchmarks, the performance of the $sym\_1.2\times$ is similar to the $sym\_1\times$ binning range, whereas the performance of the $sym\_1.5\times$ is considerably lower than the $sym\_1\times$ binning range. Therefore, $OpSparse$ adopts the $sym\_1.2\times$ binning range for the symbolic step. 

\begin{figure}[h]
\centering
\includegraphics[width=0.48\textwidth]{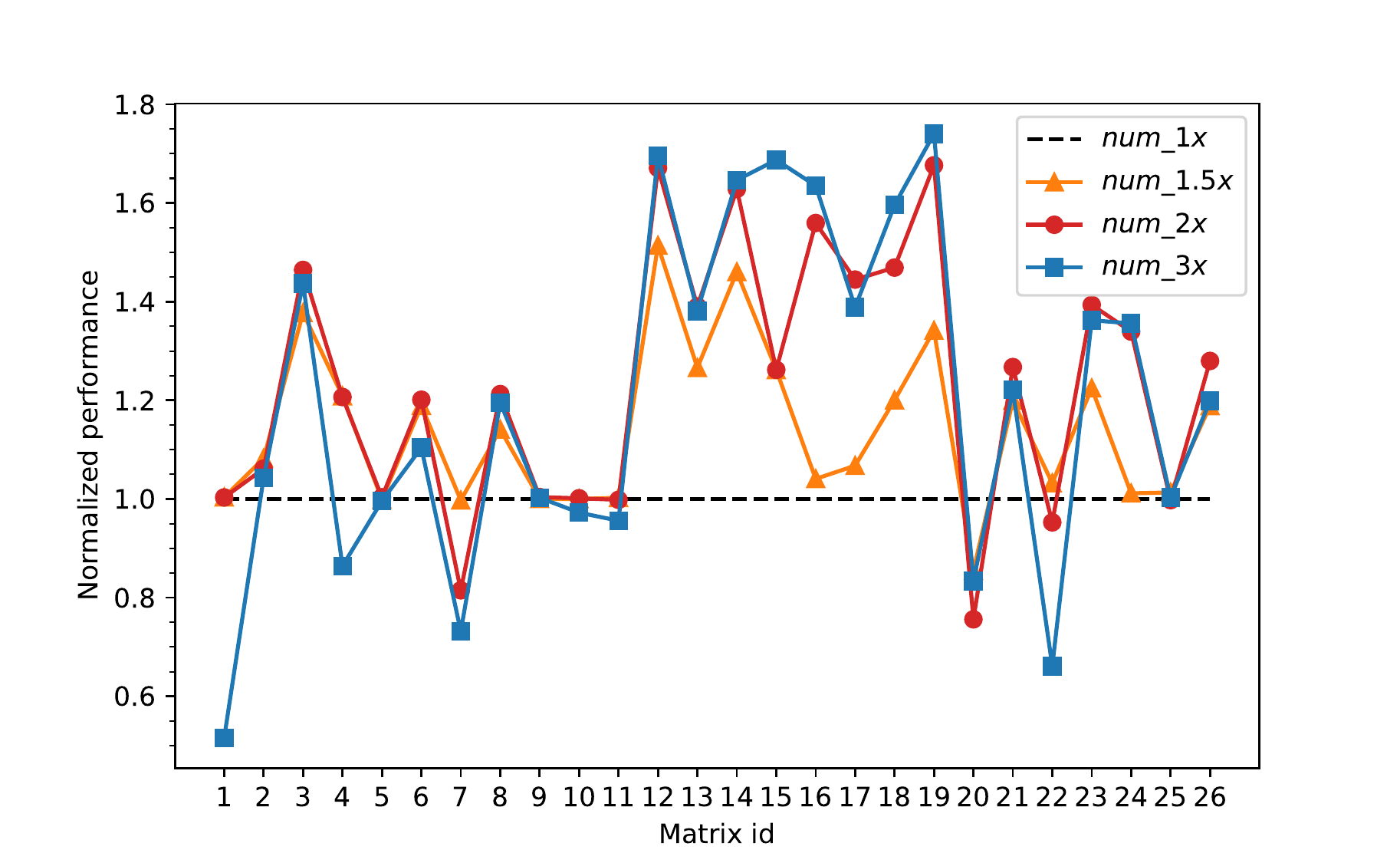}
\caption{Normalized performance of the numeric step with different binning ranges. The performance is normalized to the performance with $num\_1\times$ binning range. $num\_1\times$, $num\_1.5\times$, $num\_2\times$, and $num\_3\times$ denote the performance of the numeric step with the $num\_1\times$, $num\_1.5\times$, $num\_2\times$, and $num\_3\times$ binning ranges (shown in Table~\ref{tab:num_range}), respectively.}
\label{fig:num_range}
\end{figure}

Fig.~\ref{fig:num_range} shows the performance of the numeric step with the four kinds of binning ranges. The implementations with the $num\_1.5\times$, $num\_2\times$, and $num\_3\times$ binning ranges achieve on average $1.14\times$, $1.23\times$, and $1.20\times$ performance speedup, respectively, compared to the $num\_1\times$ binning range.
We can see that the performance is considerably improved when scaling down the binning ranges in the numeric step. We attribute the performance improvement to the lowered hash collision rate. 

Since the implementation with the $num\_2\times$ binning range achieves the average best performance than other binning ranges, our framework adopts the implementation of the $num\_2\times$ binning range for the numeric step.


\subsubsection{Load balance of the streaming multiprocessors}\label{sec:perf-SM}
When computing the \emph{webbase-1M} benchmark in our framework, one extremely large row is computed with the global memory hash table in the numeric phase, which takes one SM for execution. The computation time for the largest row is 7.6ms, whereas the computation time for all the other rows is only 20ms. This feature of the \emph{webbase-1M} benchmark shows that considering the load balance of the SMs is important. Our framework ensures that when the kernel computing the largest rows is executing, other kernels are also available for execution. As a result, the total execution time of the numeric step on \emph{webbase-1M} benchmark in our optimized framework was reduced to 21.5ms.

\subsubsection{Overlapping memory allocation with kernel execution}\label{sec:perf-overlap}
When computing the \emph{webbase-1M} benchmark in our framework,  allocating the global hash table is required in the numeric phase, which takes 1ms. Since our framework launches one kernel in the numeric phase before the memory allocation, and the execution time of the launched kernel takes more than 1ms, the execution time of the global memory allocation in our numeric step is completely hidden for the \emph{webbase-1M} benchmark.

\section{Conclusion}\label{sec:conclusion}
Efficiently performing SpGEMM on GPUs is a challenging task. We found that both high-level algorithm design and low-level architecture-specific optimizations are critical for the overall performance of SpGEMM. In this paper, we identify seven kinds of inefficient implementations in two state-of-the-art SpGEMM libraries (i.e., $nsparse$ and $spECK$). Then we propose corresponding optimizations and integrate them into a highly optimized framework $OpSparse$.
The optimizations include 1) optimizing the binning method by improving the utilization of the shared memory, 2) optimizing the hashing method by reducing the access to the hash table, 3)
improving the trade-off between hash collision rate and hardware utilization in the hashing method by setting appropriate binning ranges, 4) reducing the overheads of global memory utilization by minimizing the global memory usage of the metadata and using combined memory allocation instead of multiple separate memory allocations, 5) improving the execution parallelism by overlapping global memory allocation with kernel execution, 6) improving the load balance of the SMs in the GPU by manipulating the kernel launch orders, and 7) optimizing the kernel configuration for full occupancy. As a result, $OpSparse$ achieves on average $7.35\times$ (up to $27.8\times$), $1.43\times$ (up to $1.81\times$), and $1.52\times$ (up to $2.04\times$) speedups over three SpGEMM libraries $cuSPARSE$, $nsparse$, and $spECK$, respectively.

Moreover, We conduct comprehensive experiments to demonstrate the performance improvements of several individual optimizations. These experiments numerically show the performance improvements of our proposed optimizations as compared to the inefficient implementations in the state-of-the-art SpGEMM libraries.

\bibliographystyle{unsrt}  


\end{document}